\documentclass[longauth]{aa} 

\usepackage{graphicx}
\usepackage{natbib}
\usepackage{placeins}
\bibpunct{(}{)}{;}{a}{}{,}
\newcommand{\tess}{\emph{TESS }}
\usepackage{txfonts}
\begin{document}

\title{TESS discovery of a super-Earth and two sub-Neptunes orbiting the bright, nearby, Sun-like star HD~22946}


\author{Luca Cacciapuoti
          \inst{1,2}
          \and Laura Inno 
          \inst{3,4}
          \and Giovanni Covone
          \inst{1,4,5}
          \and Veselin B. Kostov
          \inst{6}
          \and Thomas Barclay
          \inst{6,42}
          \and Elisa V. Quintana
          \inst{6}
          \and Knicole~D.~Colon
          \inst{6}
          \and Keivan G. Stassun
          \inst{7,8}
          \and Benjamin Hord
          \inst{9}
          \and Steven Giacalone
          \inst{10}
          \and Stephen~R.~Kane
          \inst{11}
          \and Kelsey Hoffman
          \inst{12}
          \and Jason Rowe
          \inst{13}
          \and Gavin Wang
          \inst{14}
          \and Kevin I.\ Collins
          \inst{15}
          \and Karen A.\ Collins
          \inst{16}
          \and Thiam-Guan Tan
          \inst{17,39}
          \and Francesco Gallo
          \inst{1}
          \and Christian Magliano
          \inst{1}
          \and Riccardo M. Ienco
          \inst{1}
          \and Markus Rabus
          \inst{18}
          \and David~ R.~Ciardi
          \inst{19}
          \and Elise Furlan
          \inst{19}
          \and Steve~B. Howell
          \inst{21}
          \and Crystal~L. Gnilka
          \inst{21}
          \and Nicholas~J. Scott
          \inst{21}
          \and Kathryn~V. Lester
          \inst{21}
          \and Carl Ziegler
          \inst{22}
          \and C\'{e}sar Brice\~{n}o
          \inst{23}
          \and Nicholas Law
          \inst{24}
          \and Andrew W. Mann
          \inst{24}
          \and Christopher~J.~Burke
          \inst{25}
          \and Samuel~N.~Quinn
          \inst{26}
          \and Angelo Ciaramella
          \inst{3}
          \and Pasquale De Luca
          \inst{3}
          \and Stefano Fiscale
          \inst{3}
          \and Alessandra Rotundi
          \inst{3}
          \and Livia Marcellino
          \inst{3}
          \and Ardelio Galletti 
          \inst{3}
          \and Ida Bifulco
          \inst{3}
          \and Fabrizio Oliva
          \inst{27}
          \and Alton Spencer
          \inst{28}
          \and Lisa Kaltenegger
          \inst{29}
          \and Scott~McDermott
          \inst{30}
          \and Zahra~Essack
          \inst{31,32}
          \and Jon~M.~Jenkins
          \inst{21}
          \and Bill~Wohler
          \inst{12,33}
          \and Joshua N.\ Winn
          \inst{34}
          \and S. Seager
          \inst{25,35,36}
          \and Roland Vanderspek
          \inst{25}
          \and George Zhou 
          \inst{37}
          \and Avi~Shporer
          \inst{38}
          \and {Diana~Dragomir}
          \inst{40}
          \and William~Fong
          \inst{41}
          }
          
\titlerunning{A multiplanetary system around HD~22946}
\authorrunning{Cacciapuoti et al.}

\institute{Department of Physics ”Ettore Pancini”, University of Naples, Federico II
\and European Southern Observatory, Karl-Schwarzschild-Strasse 2 D-85748 Garching bei Munchen, Germany
\and Science and Technology Department, Parthenope University of Naples, CDN IC4, 80143, Naples, Italy
\and INAF-Osservatorio Astronomico di Capodimonte, Salita Moraliello, 16, 80131, Naples, Italy
\and INFN, Sezione di Napoli, Italy
\and NASA Goddard Space Flight Center, Exoplanets and Stellar Astrophysics Laboratory (Code 667), Greenbelt, MD 20771, USA
\and Vanderbilt University, Department of Physics \& Astronomy, 6301 Stevenson Center Lane, Nashville, TN 37235, USA
\and Fisk University, Department of Physics, 1000 18th Avenue N., Nashville, TN 37208, USA
\and Department of Astronomy, University of Maryland, College Park, MD 20742, USA
\and Department of Astronomy, University of California Berkeley, Berkeley, CA 94720-3411, USA
\and Department of Earth and Planetary Sciences, University of California, Riverside, CA 92521, USA
\and SETI Institute, 189 Bernardo Ave, Suite 200, Mountain View, CA 94043, USA
\and Department of Physics and Astronomy, Bishops University 2600 College St, Sherbrooke, QC J1M 1Z7, Canada
\and Tsinghua International School, Beijing 100084, China
\and George Mason University, 4400 University Drive, Fairfax, VA, 22030 USA
\and Center for Astrophysics |Harvard \& Smithsonian, 60 Garden Street, Cambridge, MA 02138, USA
\and Perth Exoplanet Survey Telescope, Perth, Western Australia
\and Departamento de Matematica y Fisica Aplicadas, Facultad de Ingegneria, Universidad Catolica de la Santisima Concepcion, Alonso de Rivera 2850, Concepcion, Chile
\and Caltech/IPAC-NASA Exoplanet Science Institute, 770 S. Wilson Avenue, Pasadena, CA 91106, USA
\and NASA Exoplanet Science Institute, Caltech/IPAC, Mail Code 100-22, 1200 E. California Blvd., Pasadena, CA 91125, USA
\and NASA Ames Research Center, Moffett Field, CA 94035, USA
\and Department of Physics, Engineering and Astronomy, Stephen F. Austin State University, 1936 North St, Nacogdoches, TX 75962, USA
\and Cerro Tololo Inter-American Observatory, Casilla 603, La Serena, Chile
\and Department of Physics and Astronomy, The University of North Carolina at Chapel Hill, Chapel Hill, NC 27599-3255, USA
\and Department of Physics and Kavli Institute for Astrophysics and Space Research, Massachusetts Institute of Technology, Cambridge, MA3902139, USA
\and Harvard-Smithsonian Center for Astrophysics, 60 Garden Street, Cambridge, MA 02138, USA
\and Institute of Space Astrophysics and Planetology INAF-IAPS, Via Fosso del Cavaliere 100, I-00133, Rome, Italy
\and Western Connecticut State University, 181 White Street, Danbury CT, 06810.
\and Department of Astronomy and Carl Sagan Institute, Cornell University, 122 Sciences Drive, Ithaca, NY 14853, USA
\and Proto-Logic LLC, 1718 Euclid Street NW, Washington, DC 20009, USA
\and Department of Earth, Atmospheric and Planetary Sciences, Massachusetts Institute of Technology, Cambridge, MA 02139, USA
\and Kavli Institute for Astrophysics and Space Research, Massachusetts Institute of Technology, Cambridge, MA 02139, USA
\and SETI Institute, Mountain View, CA 94043, USA
\and Department of Astrophysical Sciences, Princeton University, Princeton, NJ 08544, USA
\and Department of Earth, Atmospheric, and Planetary Sciences, Massachusetts Institute of Technology, Cambridge, MA 02139, USA
\and Department of Aeronautics and Astronautics, Massachusetts Institute of Technology, Cambridge, MA 02139, US
\and University of Southern Queensland, Centre for Astrophysics, West Street, Toowoomba, QLD 4350 Australia
\and Department of Physics and Kavli Institute for Astrophysics and Space Research, Massachusetts Institute of Technology, Cambridge, MA 02139, USA
\and Curtin Institute of Radio Astronomy, Curtin University, Bentley, Western Australia 6102
\and Department of Physics and Astronomy, University of New Mexico, 210 Yale Blvd NE, Albuquerque, NM 87106, USA
\and Department of Physics and Kavli Institute for Astrophysics and Space Research, Massachusetts Institute of Technology, Cambridge, MA 02139, USA
\and University of Maryland, Baltimore County, 1000 Hilltop Circle, Baltimore, MD 21250, USA
}
   
\authorrunning{Cacciapuoti et al.}

\abstract{We report the Transiting Exoplanet Survey Satellite (TESS) discovery of a three-planet system around the bright Sun-like star HD~22946 (V$\approx$8.3 mag), also known as TIC~100990000, located 63 parsecs away.
The system was observed by TESS in Sectors 3, 4, 30 and 31 and two planet candidates, labelled TESS Objects of Interest (TOIs) 411.01 (planet $c$) and 411.02 (planet $b$), were identified on orbits of 9.57 and 4.04 days, respectively. In this work, we validate the two planets and recover an additional single transit-like signal in the light curve, which suggests the presence of a third transiting planet with a longer period of about 46 days.
We assess the veracity of the TESS transit signals and use follow-up imaging and time series photometry to rule out false positive scenarios, including unresolved binary systems, nearby eclipsing binaries or background/foreground stars contaminating the light curves. 
Parallax measurements from Gaia Early Data Release 3, together with broad-band photometry and spectroscopic follow-up by TFOP allowed us to constrain the stellar parameters of TOI-411, including its radius of $1.157 \pm 0.025 R_\odot $. 
Adopting this value, we determined the radii for the three exoplanet candidates and found that planet $b$  is a super-Earth, with  a radius of $1.72 \pm 0.10$ $R_\oplus$, while planet $c$ and $d$ are sub-Neptunian planets, with radii of  $2.74 \pm 0.14$ $R_\oplus$ and $3.23 \pm 0.19$ $R_\oplus$ respectively.
By using dynamical simulations, we assessed the stability of the system and evaluated the possibility of the presence of other undetected, non-transiting planets by investigating its dynamical packing. We find that the system is dynamically stable and potentially unpacked, with enough space to host at least one more planet between $c$ and $d$.
Finally, given that the star is bright and nearby, we discuss possibilities for detailed mass characterization of its surrounding worlds and opportunities for the detection of their atmospheres with the \textit{James Webb Space Telescope.}}


\maketitle
%
\section{Introduction}

Multi-planetary systems are crucial cosmic environments to probe our theories of planet formation and evolution.
Exoplanets surveys have revealed an unexpected diversity of such systems, including exoplanets for which no analog is present in the Solar System and planetary systems whose orbital architecture is very different from the one observed in our own \citep[see e.g.,][]{Gillon2017, Naef2001, Winn2011, Miret-Roig2021}. 
Transiting planetary systems around nearby, bright stars offer the opportunity for in-depth observational studies to characterize different planets formed in a common environment.
The Transiting Exoplanet Survey Satellite \citep[\tess ;][]{Ricker2015} is a space-borne NASA mission launched in 2018 to survey the sky for transiting exoplanets around these stars. 
To date, it has contributed to the discovery of more than 200 exoplanets in the Galaxy, with more than 5700 candidates on hold for confirmation\footnote{See the NASA Exoplanet Archive
at \url{https://exoplanetarchive.ipac.caltech.edu/}}. 
These discoveries include planets of different sizes, ranging from Mars-like bodies \citep[e.g., L98-59 b,][]{Kostov2019} to gas giants larger than Jupiter \citep[e.g., TOI-640 b,][]{Rodriguez2021}, with the majority of them being super-Earths ($1.25 < R_{\oplus} < 1.75$) and sub-Neptunes ($1.75 < R_{\oplus} < 3.5$), according to the classification criteria defined by \citet{Fulton2017}. 

Exoplanets discovered around bright stars are ideal targets for detailed characterization, including the measurements of their mass via Doppler spectroscopy and the determination of their atmospheric proprieties through transmission or emission spectroscopy. 
Besides, multiplanetary systems offer the additional opportunity to perform comparative exoplanetology studies, since we can constrain the formation and evolution of a group of planets that share a common origin, given the fixed proprieties of the host star. 
In fact, for these systems, we can gather enough information to constrain the orbits and dynamics of the system, and hence to infer the underlying formation and evolution scenarios, see e.g., \citet{Ragozzine2010,Lissauer2011,Fabrycky2014}.

In this work, we report the TESS discovery of a multi-planetary system consisting of a Super-Earth and two sub-Neptune-sized planets orbiting the bright (V$\approx$8.3 mag), nearby (63~pc), Sun-like star HD~22946, indicated in the TESS Input Catalog \citep{Stassun2018} as TIC 100990000 and as TOI 411 in the TESS Objects of Interest list on ExoFOP-TESS\footnote{https://exofop.ipac.caltech.edu/tess/}. 

Sub-Neptunes and super-Earths are of particular interest for planetology studies because, while they represent the highest fraction of planets detected in the Galaxy, they are not observed in the Solar System. 
The detection and in-depth characterization of such planets orbiting Sun-like stars can help us constrain the formation and evolution pathways along which they form and gain a better understanding of different planetary systems \citep{Bitsch2019}.
For example, a depression in the distribution of exoplanets radii between 1.5 and 2 $R_\oplus$ \citep{Fulton2017,Owens2017} seems to indicate that these exoplanets accreted an atmosphere at the time of formation that progressively thinned out due to heat-driven processes, such as stellar radiation photo-evaporation \citep{Fulton2018} or core-powered mass loss \citep{Ginzburg2018}. Moreover, \citet{Loyd2020} found that doubling the known sample of exoplanets with radii $R < 4 R_\oplus$ is required to confirm or exclude the role of photo-evaporation in the supposed atmospheric loss of these worlds.
Thus, by studying planets in this range, we can shed light on the physical processes that can lead to the loss of a primary atmosphere and the possible accretion of a heavier secondary one.

The paper is organized as follows. In Section~\ref{star} we characterize the host star, in Sect.~\ref{obs} we present the detection of the three transiting candidates from TESS observations and their characterization. 
In Sect.~\ref{NEB} we analyze the follow-up data to validate their planetary nature. 
In Sect.~\ref{syst}, we explore the dynamics of the system and the potential for atmospheric characterization of its planets. Finally, in Sect.~\ref{conc}, we summarize our findings.

\section{Characterization of the host star}
\label{star}

Characterizing the physical proprieties of the host star, such as its mass, $M_{\star}$,  radius, $R_{\star}$ and effective temperature $T_{\rm{eff}}$, is propaedeutic to constraining the corresponding properties of its planetary system. 
Therefore, we first focus on determining these quantities for HD~22946, together with its spectral class, surface gravity, $\log g$,  metallicity, [Fe/H], and sky-projected rotational velocity, $v \sin i$.

HD~22946 is a bright ($V\approx$8.3 mag) main-sequence star located in the Southern Hemisphere, characterized by a high proper motion and some photometric variability. Basic stellar information is given in \ref{tab:stellardata}.

\subsection{Stellar variability} 
In order to understand better the photometric variability of the star, we searched for publicly-available light curves of HD~22946, we only found multi-epoch photometry from the Hipparcos Epoch Photometry catalog \citep{Hipparcos},which includes a hundred measurements over about 3 years, with 3 to 10 measurements per night each month. \footnote{Light curves for this star are available in ASAS-SN Sky Patrol database \citep[][and references therein]{Shappee2014,Jayasinghe2021} and in ASAS \citep{asas}, but the stellar magnitude is close to the surveys' saturation limit so we discarded this data. Epoch-photometry from Gaia DR3 is not available for this star.}

\begin{figure}
    \centering
    \includegraphics[width=\linewidth]{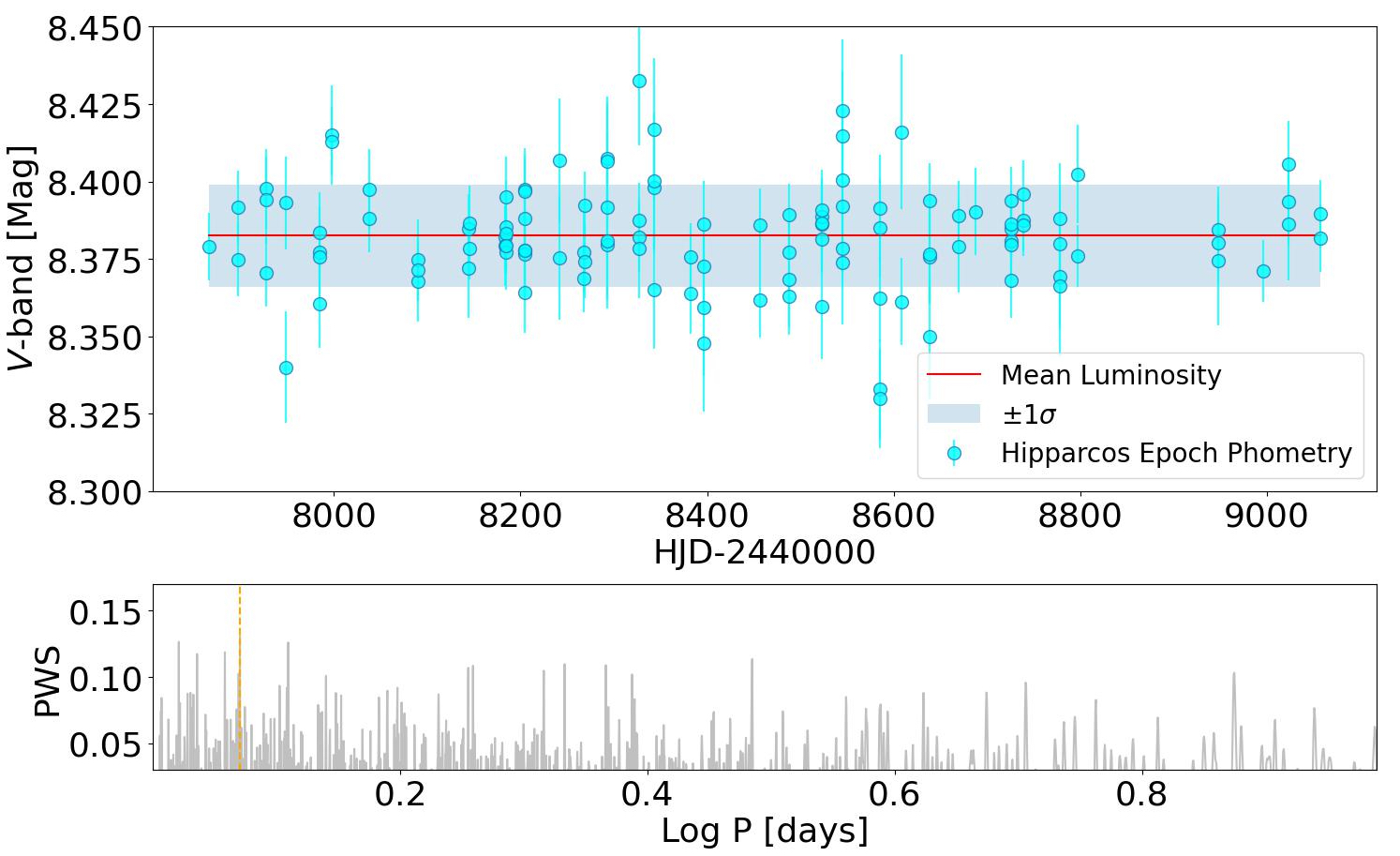}
    \caption{Light curve of HD~22946 from the Hipparcos Epoch Photometry data ($V-$ band, top panel), and the corresponding power spectrum (bottom panel). In the top panel, the mean luminosity and standard deviation are also shown: the scatter in the photometric data is compatible with a low-amplitude variability.
    The peak of the power spectrum, indicated by the orange vertical line, is found at $\approx$0.073 days, and it is likely due to instrumental aliasing.}
    \label{fig:hipp}
\end{figure}

The light curve from the Hipparcos data shows a small amplitude variation, with a maximum-minimum difference of about $\approx$ 0.1 mag, or six times the standard deviation $\sigma$ of the measurements, as can be seen in Figure~\ref{fig:hipp}, where we also show the computed Lomb-Scargle \citep{Lomb1976,Scargle1982} periodogram using a Nyquist factor of 10 following recommendation from \cite{Richards2012} and \cite{VanderPlas2018}. The peak of the power spectrum is found at around 0.073 days, close to the 0.076 days data gaps of Hipparcos observations \citep{Percy2002}, therefore it is likely due to aliasing. Indeed, the star is classified as static (class "C") in the Hipparcos catalog.

The TESS light curve of HD~22946 also shows mild signs of variability, especially in sectors 30 and 31, as we further discuss in the next section.

\begin{table*}
\caption{Summary of stellar data used in this work.}	
\label{tab:stellardata}
\centering
\begin{tabular}{lcc}
\hline\hline
Parameters & Value & Source \\ \\
\hline \\
Name & TOI-411 & \citet{Guerrero2021}\\
TIC ID & 100990000 & TICv8.1\\
Alt. name & HD~22946 & \\ \\
\multicolumn{3}{c}{\em Astrometric Properties} \\
$\alpha$ R.A. (hh:mm:ss) & 03:39:16.761  & Gaia EDR3 \\
$\delta$ Dec. (dd:mm:ss) & -42:45:45.185  & Gaia EDR3 \\
$\mu_{\alpha}$ (mas~yr$^{-1}$) & $-51.618 \pm 0.015$ & Gaia EDR3 \\
$\mu_{\delta}$ (mas~yr$^{-1}$) & $-110.546 \pm 0.021$  & Gaia EDR3 \\
Barycentric RV (km~s$^{-1}$)  & $16.91 \pm 0.15$ & Gaia DR2 \\
Distance (pc) & $62.87 \pm 0.05$ & Gaia EDR3\\ \\ 
\multicolumn{3}{c}{\em Photometric Data} \\
TESS (mag)\dotfill& $7.757 \pm 0.006$ & TIC v8.1 \\
FUV (mag)\dotfill& $19.45 \pm 0.16$ & GALEX GR6\\
NUV (mag)\dotfill& $13.014 \pm 0.005$ & GALEX GR6\\
$B_T$ (mag)\dotfill& $8.903 \pm 0.017$ & Tycho-2\\
$V_T$ (mag)\dotfill& $8.318 \pm 0.012$ & Tycho-2\\
$G$-band (mag)\dotfill& $8.1281 \pm 0.0003$ & Gaia DR2  \\
$G_{BP}$-band (mag)\dotfill& $8.4206 \pm 0.0003$ & Gaia DR2  \\
$G_{RP}$-band (mag)\dotfill& $7.7173 \pm 0.0003$ & Gaia DR2  \\
$B$ (mag)\dotfill& $8.79 \pm 0.03 $ & TIC v8.1  \\
$V$ (mag)\dotfill& $8.27 \pm 0.03 $ & TIC v8.1  \\
$J$ (mag)\dotfill& $7.25 \pm 0.04 $ &  2MASS \\
$H$ (mag)\dotfill& $7.04 \pm 0.04 $ &  2MASS \\
$K_{\rm{s}}$ (mag)\dotfill& $6.98 \pm 0.03  $ & 2MASS \\
$w_1$ (mag)\dotfill& $6.91 \pm 0.04 $ &  WISE \\
$w_2$ (mag)\dotfill& $6.93 \pm 0.02 $ &  WISE \\
$w_3$ (mag)\dotfill& $6.95 \pm 0.02  $ & WISE \\
$w_4$ (mag)\dotfill& $6.86 \pm 0.05  $ & WISE \\ \\ 
\multicolumn{3}{c}{\em Spectroscopic Data in TFOP} \\ 
UT 2019-02-09  & SNR=34 &  NRES@LCO\\
UT 2019-02-19  & SNR=40 &  NRES@LCO\\
UT 2019-02-23  & SNR=40 &  NRES@LCO\\
UT 2019-02-27  & SNR=41 &  NRES@LCO\\
UT 2019-02-23  & SNR=86 &  CHIRON@SMARTS\\
UT 2019-08-09  & SNR=70 &  CHIRON@SMARTS \\
UT 2021-01-10  & SNR=63 &  CHIRON@SMARTS   \\
UT 2021-08-28  & SNR=50 &  CHIRON@SMARTS \\
\hline 
\end{tabular} 
\end{table*}

\begin{figure}
    \centering
    \includegraphics[width=\linewidth,trim=210 140 175 150,clip]{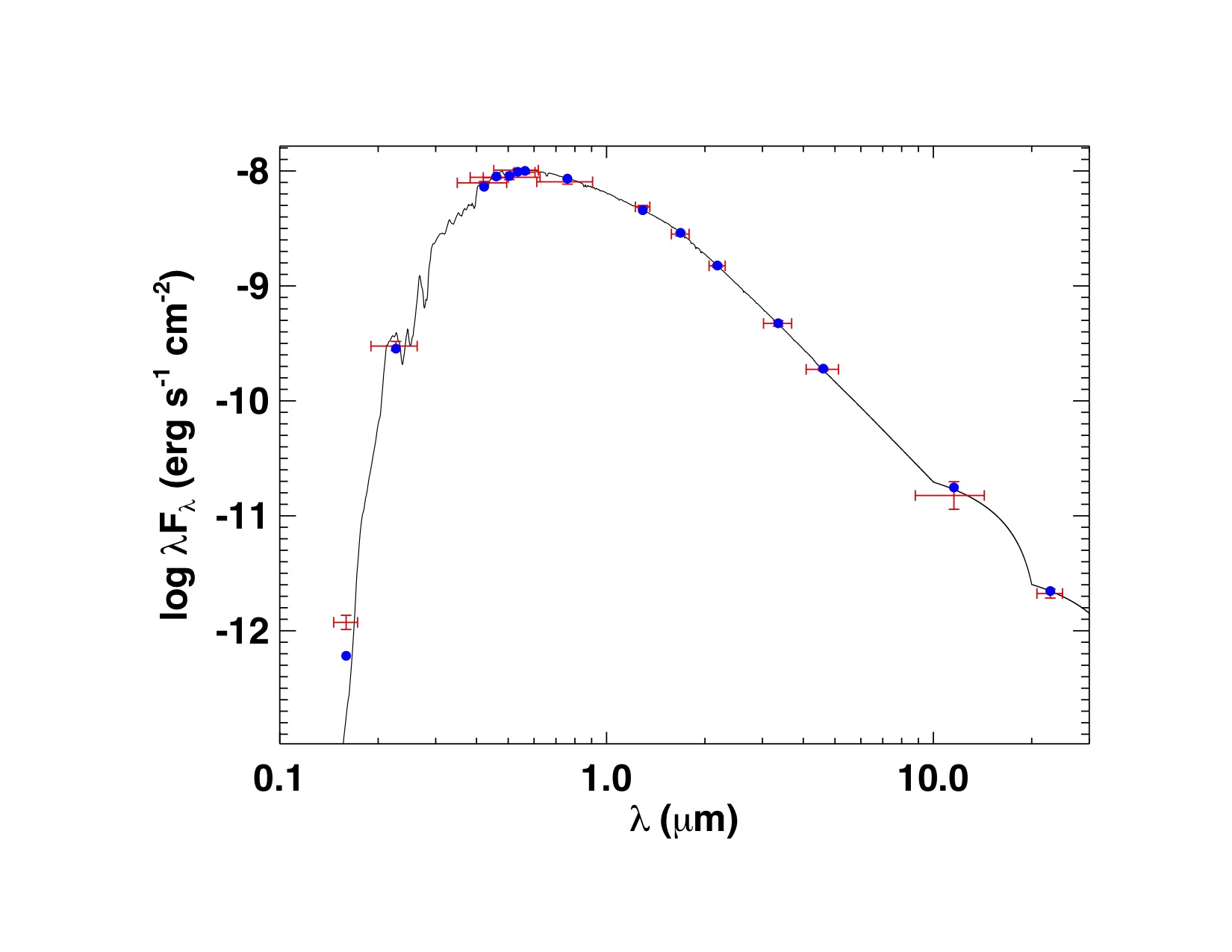}
    \caption{Spectral energy distribution of HD~22946. The blue symbols and their vertical error bars represent the photometric measurements available for this source and listed in Table~\ref{tab:stellardata}. The horizontal bars indicate instead the width of the passband for each data point. The overplotted black line is the best fit model, which allows us to derive the stellar parameters.}
    \label{fig:sed}
\end{figure}

\subsection{Stellar parameters} 
\label{section:stellar_param}
Light curves from photometric transits only allow the measurement of the planet-to-star radius ratio, hence we need to accurately determine the stellar radius in order to infer the radii of the transiting planets. 

In the literature there are several estimates for the mass and radius of HD~22946:
i) the values from Gaia DR2 \citep{gaia_dr2}, and ii) the values obtained by \citet[][hereinafter K19]{Kervella2019} by using the $V, V-K$ surface brightness relations for the radius \citep{Kervella2004} and the isochrone fitting method for the mass \citep{Girardi2000}.
Both these approaches depend on the measured parallax.
Besides these estimates, we can use two different and independent methods to infer the radius of the star and the other stellar parameters. 
The first method relies on the analysis of the spectra collected through the TESS Followup Observing Program (TFOP\footnote{\url{https://tess.mit.edu/followup/}}, \citep{collins:2019}): high-resolution, high signal-to-noise (SNR) optical spectra of HD~22946 were taken at 4 different epochs
with both the Network of Robotic Echelle Spectrographs (NRES) \citep{Siverd2018} of the Las Cumbres Observatory (LCO; \citealt{Brown2013}) \citep{Siverd2016} and the CHIRON spectrograph \citep{tokovinin18} on the 1.5m Small and Moderate Aperture Research Telescope System (SMARTS) at the Cerro Tololo Inter-American Observatory (CTIO).

The observations are detailed in Table~\ref{tab:stellardata}.
NRES data cover a 380-860 nm spectral band with a resolution of 53,000 and were processed by the dedicated data reduction pipeline \citep{McCully2018}, and then analyzed by using \textsf{SpecMatch} \citep{Petigura2017}. 
CHIRON spectra cover the 450-890 nm range with a resolution of 80000. The spectra were collected by the instrument team and reduced following the optimal extraction method described in \citet{Paredes2021}. We derived radial velocities by fitting the line profiles of the spectra, extracted via Least Squares Deconvolution (LSD) of the observed spectrum against synthetic templates \citep{Donati1997}, and we estimated the spectroscopic stellar parameters by using the gradient-boosting regressor implemented in the \texttt{scikit-learn} python module, previously trained on spectra classified by \texttt{SPC} \citep{Buchhave2012}. Our analysis of the line profiles reveals negligible rotational broadening and no evidence of double lines, suggesting that the star is inactive and not host to unresolved, close-in, stellar companions.

In Table~\ref{tab:stellarparameters} we list the radial velocity $V_{\rm{rad}}$ measured at each epoch, and the weighted average with associated uncertainties for all the stellar parameters obtained by the spectral analysis, with the prior of the Gaia parallax.  

We note that the $V_{\rm{rad}}$ measured at the different epochs from the same spectrograph are all consistent with each other. A zero-point discrepancy among the measures obtained from the two instruments is expected.
NRES observations do not show significant variation at the level of 600 m/s over 18 days while CHIRON spectra do not show significant variation at the level of 100 m/s over 2.5 years. Moreover, the $V_{\rm{rad}}$ from Gaia DR2 also shows a small uncertainty, i.e. a small $rms$ from the different epoch observations, and is consistent with the value measured from the NRES spectra. All the above evidence further confirm that the star does not have a bound companion. 

The second method used to obtain stellar parameters is based on the fitting of the spectral energy distribution (SED), i.e., the flux emitted by the star across a broad wavelength range, which yields a semi-empirical determination of the stellar radius, mass, and age. 

We used the photometric data listed in Table~\ref{tab:stellardata}, following the method described in \citet{Stassun2016} and \citet{Stassun2018}.
We retrieved public photometric data in different passbands as follows:

i) FUV and NUV magnitudes from the GALEX catalog; ii) $BV$ magnitudes from the catalog of \citet{Mermilliod:2006}; iii) $B_T V_T$ magnitudes from the Tycho-2 catalog; iv) $G G_{\rm BP} G_{\rm RP}$ magnitudes from Gaia EDR3; v) $JHK_S$ magnitudes from the 2MASS catalog \citep{Skrutskie2006,Cutri2003}; and vi) magnitudes in the mid-infrared bands at 3.4, 4.6, 12 and 22 $\mu$m from the WISE catalog \citep{Wright2010}.
The full wavelength range covered by the data goes from 0.15 to 22$\mu$m, as shown in Figure~\ref{fig:sed}.

We performed a fit using the Kurucz stellar atmosphere models, adopting the spectroscopically determined $T_{\rm eff}$, $\log g$, and [Fe/H], as well as the $v\sin i$ from \citet{Pribulla2014}. 
The remaining free parameter is the interstellar extinction, $A_V$, which we limited to the maximum line-of-sight value from the dust maps of \citet{Schlegel:1998}. The fit is reasonably good, with a reduced $\chi^2$ of 1.9 and best fit $A_V$=0.02$\pm$0.01. Integrating the (unreddened) SED gives $F_{\rm bol}$=1.289$\pm$0.030 $10^{-8}$ erg~s$^{-1}$~cm$^{-1}$, that, combined with the Gaia EDR3 parallax \citep[with no offset applied; see, e.g.,][]{StassunTorres:2021} gives $R_{\star}$=1.139$\pm$0.035$R_{\odot}$.
With the spectroscopic $\log g$, we then calculate a value of $M_{\star}$=1.13$\pm$0.12$M_{\odot}$, which is consistent with that determined by adopting the \citet{Torres2010} empirical relations, namely 
$M_{\star} =1.12 \, \pm \, 0.06 \, M_{\odot}$. 
Finally, we can estimate the stellar rotation period via the observed $R'_{\rm HK}$ 
activity index \citep{Murgas2013} 
and the empirical rotation-activity relations of \citet{Mamajek:2008}, which yields $P_{\rm rot} = 10.6\pm0.9$~days .
Using these rotation and activity measures together with the activity-age relations of \citet{Mamajek:2008} gives an estimated stellar age of $\tau_\star = 5 \pm 1$~Gyr.

The derived stellar parameters from both methods are consistent with each other and listed in Table~\ref{tab:stellarparameters}, together with the best values adopted through the next steps of the analysis, that were obtained by performing the error-weighted average of all the available parameters. 

Summarising, we found that HD~22946 is a late-type F main-sequence star, with $T_{\rm{eff}}$=6040$\pm$48 K,  $\log g$=4.26$\pm$0.15 dex, [Fe/H]=-0.14 $\pm$ 0.07 dex,   $M_{\star}$=1.104$\pm$0.012 $M_{\odot}$ and $R_{\star}$= 1.157$\pm$0.025 $R_{\odot}$.

\begin{table*}
\centering 
\caption{Stellar data}
\label{tab:stellarparameters}
\begin{tabular}{ccc}
\hline\hline
Parameter & Value & Source \\ \\
\hline \\
& Literature Data & \\
$T_{\rm{eff}}$ (K) & 6115 $\pm$ 324 & Gaia DR2\\
$A_G$ (mag) & 0.05 $\pm$ 0.5 & Gaia DR2 \\
$V_{\rm{rad}}$  (km s$^{-1}$) & 16.91 $\pm$0.15 & Gaia DR2 \\
$R_{\star}$ ($R_{\odot}$)& 1.14$\pm$ 0.11 & Gaia DR2\\ 
$M_{\star}$ ($M_{\odot}$)& 1.158 $\pm$ 0.058 & K19\\ 
$R_{\star}$ ($R_{\odot}$)& 1.115 $\pm$  0.056 &   K19\\ 
$V_{\rm{rad}}$  (km s$^{-1}$) & 16.15 $\pm$0.04 & \citet{Pribulla2014} \\
$v \sin i$  (km s$^{-1}$) & 7.05$\pm$0.2 &  \citet{Pribulla2014} \\ 
Spectral Type  & F7/F8V &  \citet{Pribulla2014} \\ \\
& High-resolution spectroscopy & \\
$V_{\rm{rad}}$  (km s$^{-1}$) & 16.6 $\pm$0.3  & NRES@LCO \\
$V_{\rm{rad}}$  (km s$^{-1}$) & 16.2 $\pm$0.3  & NRES@LCO \\
$V_{\rm{rad}}$  (km s$^{-1}$) & 16.5 $\pm$0.8  & NRES@LCO \\
$V_{\rm{rad}}$  (km s$^{-1}$) & 16.8 $\pm$0.4  & NRES@LCO \\
$V_{\rm{rad}}$  (km s$^{-1}$) & 17.589 $\pm$ 0.033 & CHIRON@SMARTS \\
$V_{\rm{rad}}$  (km s$^{-1}$) & 17.624 $\pm$ 0.022  & CHIRON@SMARTS \\ 
$V_{\rm{rad}}$  (km s$^{-1}$) & 17.685 $\pm$ 0.038  & CHIRON@SMARTS \\ 
$V_{\rm{rad}}$  (km s$^{-1}$) & 17.620 $\pm$ 0.037 & CHIRON@SMARTS \\ \\
$T_{\rm{eff}}$ (K) & 6210 $\pm^{173}_{158}$ & This work, NRES\\
$\log g$ & 4.4 $\pm$ 0.2  & This work, NRES \\
$[Fe/H]$ & -0.05 $\pm^{0.08}_{-0.05}$ & This work, NRES\\ 
$v \sin i$ (km s$^{-1}$) & 3.4 $\pm$ 0.9  & This work, NRES\\ 
$M_{\star}$ & 1.104 $\pm$ 0.012  & This work, NRES\\ 
$R_{\star}$ & 1.157 $\pm$ 0.025 & This work, NRES\\ 
$T_{\rm{eff}}$ (K) & 6026 $\pm$ 50 & This work, CHIRON\\
$\log g$ & 4.22 $\pm$ 0.10  & This work, CHIRON \\
$[Fe/H]$ & -0.22 $\pm$ 0.08 & This work, CHIRON\\ 
$v \sin i$ (km s$^{-1}$) & 3.2 $\pm$ 0.5  & This work, CHIRON\\ 
\\
&SED fitting from broad-band photometry& \\
$A_V$ & 0.02 $\pm$ - 0.01  & This work\\ 
F$_{bol}$  (erg s$^{-1}$ cm$^{-2}$) & 1.289 $\pm$ 0.030 & This work\\
$R_{\star}$ ($R_{\odot}$)  & 1.139 $\pm$ 0.035 & This work\\ 
$M_{\star}$  ($M_{\odot}$) & 1.12 $\pm$ 0.06 & from $\log g$ and $R_{\star}$\\ 
$M_{\star}$ ($M_{\odot}$) &  1.13 $\pm$ 0.12 & from \citet{Torres2010}'s empirical relations\\ 
Age (Gyr) &  5 $\pm$ 1  &  This work\\
\\
&Best values adopted through this work& \\
$T_{\rm{eff}}$ (K) & 6040 $\pm 48$ & This work \\
$\log g$ & 4.26$\pm$  0.15 & This work  \\
$[Fe/H]$ & -0.14$\pm$  0.07 & This work \\ 
$R_{\star}$ ($R_{\odot}$)  & 1.157 $\pm$ 0.025 & This work\\ 
$M_{\star}$  ($M_{\odot}$) & 1.104 $\pm$ 0.012 & This work\\ 
\hline 
\end{tabular} 
\end{table*}
\section{Discovery and characterization of HD~22946 planets}
\label{obs}

\begin{figure}
    \centering
    \includegraphics[width=\linewidth]{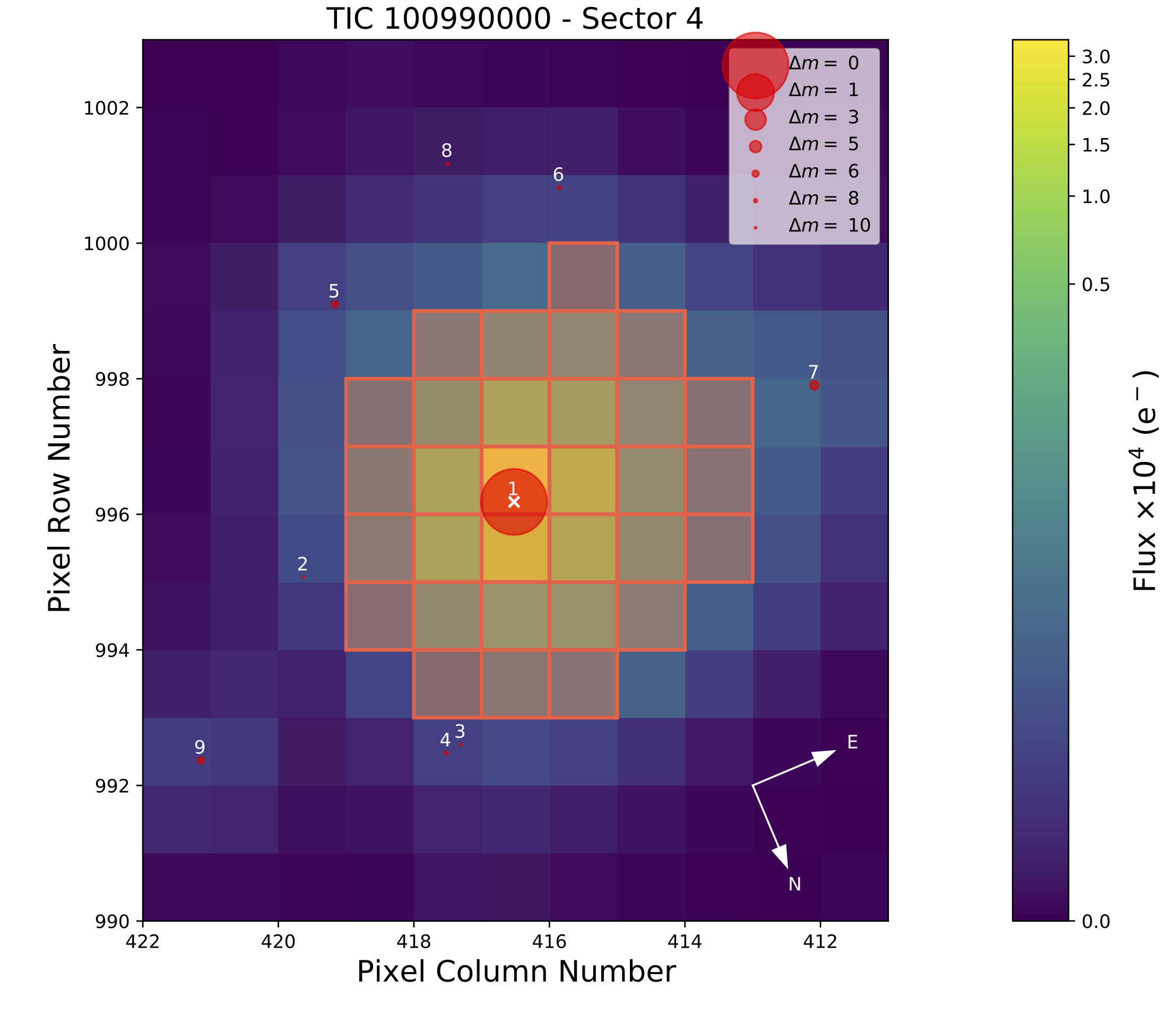}
    \caption{TESS image of HD~22946 taken during Sector 4 observations. The target is in the center, labeled with a "1" and a white cross. Nearby sources down to +10 mag are shown as smaller red dots and labeled in order of distance from the target. The SPOC-based, optimal aperture for the specific sector is overplotted in red. The color scheme represents the flux scale on the pixels. This image has been created using the \textsf{tpfplotter} python package \citep{Lillo-Box2020}.}
    \label{fig:tpf}
\end{figure}

In this Section we present the TESS observations of HD~22946 and the analysis performed to detect and characterize the transit signals in its light curve. 

\subsection{TESS observations}
\begin{figure*}
    \centering
    \includegraphics[width=\linewidth]{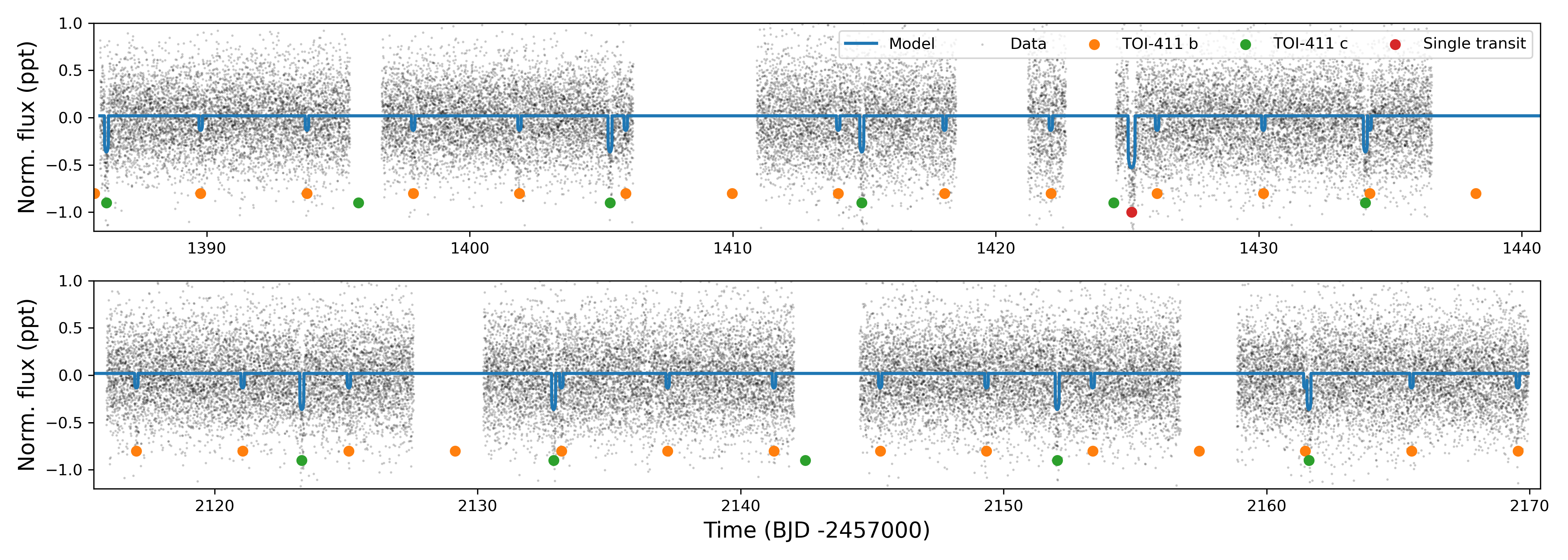}
    \caption{Full light curve of TESS sectors 3, 4 (top panel) and 30, 31 (bottom panel). The blue line is the best fit model. The transits of planet \textsf{b} and \textsf{c} are highlighted with orange and green dots, respectively. The single transit event is highlighted with a red dot.}
    \label{fig:lctot}
\end{figure*}

TESS observed HD~22946 
with Camera 3 in sectors 3 and 4 (2018-Sep-20 to 2018-Oct-18 and 2018-Oct-18 to 2018-Nov-15) during its primary mission and sectors 30 and 31 (2020-Sep-22 to 2020-Oct-21 and  2020-Oct-21 to 2020-Nov-19) during the extended mission. Fig.\ref{fig:tpf} displays a single TESS frame of HD~22946 taken during Sector 4 observations.
The first planet search has been carried out with the Science Processing Operation Center (SPOC) pipeline \citep{Jenkins2016}. 
The SPOC pipeline is designed to identify potential exoplanet transits using a wavelet-based, adaptive noise-compensating matched ﬁlter with the Transiting Planet Search code \citep{Jenkins2002}. 
Subsequently, it fits a transit model to the detected signals \citep{Li2019} and automatically performs a suite of tests to vet their planetary nature \citep{Twicken:DVdiagnostics2018}. These are then collected in a Data Validation Report (DVR), available on both the ExoFOP-TESS and MAST\footnote{https://mast.stsci.edu/} archives. 
At this point, the pipeline masks the detected signals and runs from the top to identify new signatures. 
It stops after eight tries or in case the latest found signal is below a certain SNR threshold. 
Finally, the pipeline generates two different light curves for the detected TOIs: the first one is extracted through simple-aperture photometry (SAP-FLUX; \citet{Twicken2010}; \citet{Morris2020}), whilst the other one has an additional pre-search data conditioning procedure (PDCSAP-FLUX; \citet{Stumpe2014}). 
This final step removes long-term trends and contamination due to nearby stars. 
Two planet candidates have been discovered during this preliminary analysis and added as TESS TOI-411.01 and 411.02 to ExoFOP-TESS. 
These candidates have periods of 9.57 days and 4.04 days, with transit epochs at 1385.72367 and 1386.189174 Barycentric TESS Julian Date (BTJD\footnote{BTJD = BJD - 2457000.0}), respectively. 
The full TESS light curve is shown in Fig.~\ref{fig:lctot}. 
The transits of the candidates are highlighted. 

\subsection{Light curve analysis}
\label{lc}
We downloaded and worked on the PDCSAP light curve for TOI~411 using the \textsf{lightkurve} package \citep{Lightkurve2018}. This code makes use of an automatically-generated optimal aperture but we checked that different tentative choices of apertures around this target do not affect the depth of the transit signals. Starting with this time series, we first recover the candidate events and subsequently fit them to retrieve physical and orbital parameters.

\subsection{Candidates detection}
\label{sec:detection}
We used the Transit Least Squares (TLS) package \citep{Hippke2019} to recover the candidate signals. This code allows searching for periodic, transit-like signals in a light curve by brute-force fitting templates of transits built over a grid of parameters. We decided to adopt this code with respect to the more commonly used Box Least Squares (BLS, \citet{kovacs2002}) algorithm, because 
it also accounts for the effect of stellar limb-darkening on the shape of transits, and, according to the authors, this improvement yields a $\sim$10\% higher detection efficiency (and similarly lower false alarm rates) 
compared to BLS, which is especially relevant for low SNR signals. 
The brute-force fitting consists of the phase-folding of the light curve over a grid of trial periods, transit epochs and duration, and computing the $\chi^2$ at each solution. 
The solution is then found by minimizing the $\chi^2$. 
The period grid is a uniformly-spaced set of $N$ possible values within a given range. The upper limit of the range is set by the minimum number of transits we want to find over the entire observation window, which we set to two. 
The lower limit is such that the candidate orbits just wide of the Roche limit of its star (in the assumption of a planet density of $\rho_{p} = 1 \, {\rm g} \, {\rm cm^{-3}}$). In the case of HD~22946, this limit corresponds to 0.6 days.
We performed the first TLS run independently on the couples of sectors [3,4] and [30,31] to ease the computational burden of the search. 
The first run on sectors 3 and 4 yields a periodic event with a period of $\sim9.57$ days, which is consistent with the ExoFOP TESS listed period for TOI~411.01.
Masking this signal and running a second search, 
we recover the events associated with TOI~411.02 on a $\sim4.04$ days period. A third TLS run does not return consistent periodic events. The same results are found for sectors 30 and 31. 
In addition, by visually inspecting the available sectors, we notice two additional events: a transit-like feature in sector 4 at BTJD $\sim1425$ and another dip in sector 30 at BTJD $\sim2136.5$.  However, the latter event presents an asymmetric shape and it overlaps the momentum dump reported in TESS Data Release Notes no.45 (DRN45, and Fig.4 and Fig.7 therein) on MAST\footnote{ \url{https://archive.stsci.edu/tess/tess_drn.html}} at the same epoch. 
The former event, instead, does not correspond to any reported spacecraft maneuver epoch. 
Hence, we only consider the sector 4 transit-like signal hereafter. This newly found signal is then a single event candidate exoplanet whose period is yet to be defined (see Fig.~\ref{fig:lctot} of this work).

This single event candidate was also recently reported independently on ExoFOP-TESS (as TIC 100990000.03) by A.S., a member of the Planet Hunters TESS citizen science project \citep{Eisner2021}. 

We note that TESS is not scheduled to observe this target again during its upcoming extended mission and therefore follow-up observations with other facilities will be needed to confirm this newly found candidate exoplanet.

\subsection{Transit modeling}
\label{modeling}
\begin{figure}
   \centering
   \includegraphics[width=\linewidth]{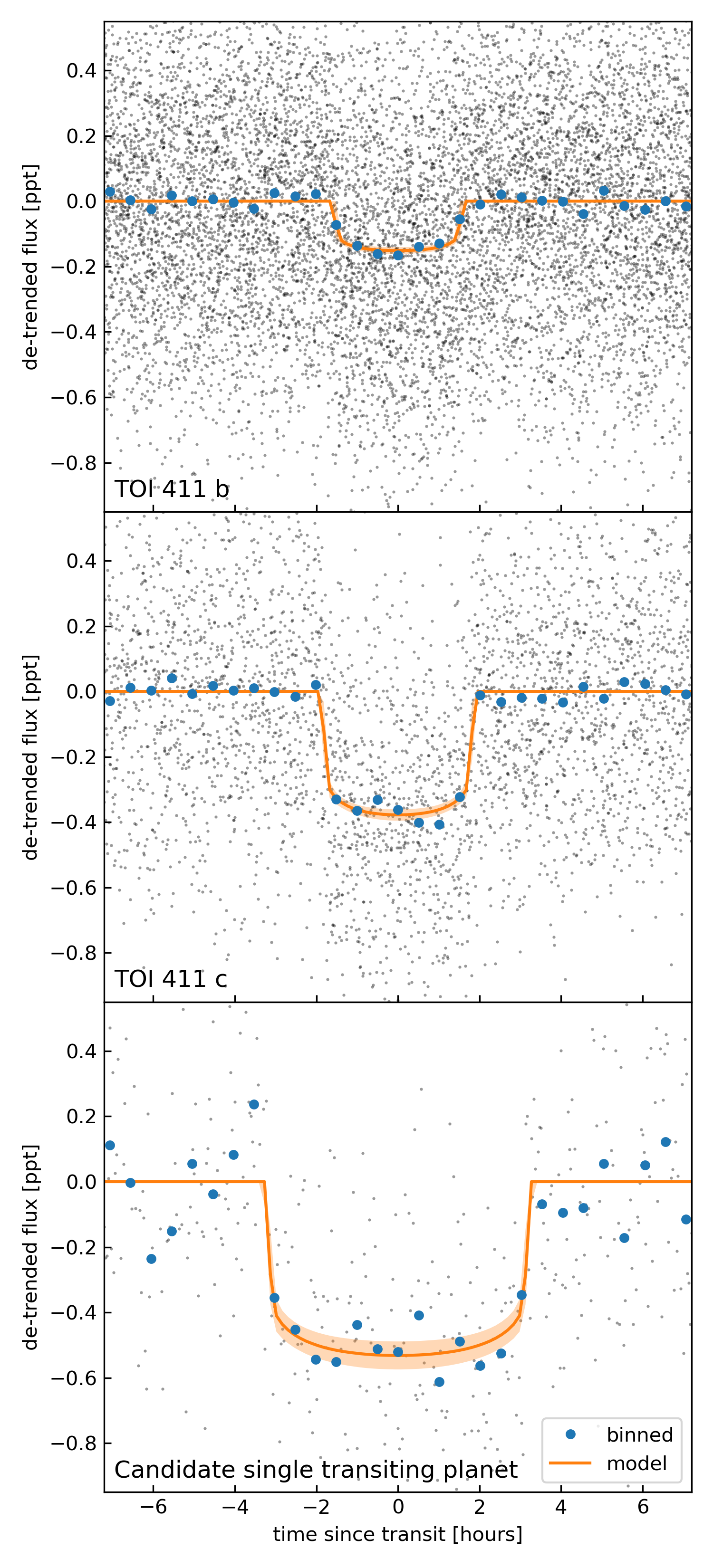}
   \caption{Folded TESS light curves from sectors 3, 4, 30 and 31 corresponding to the transits of planet $b$ (top panel), planet $c$ (middle panel) and the single event transit of candidate planet $d$ from sector 4 (bottom panel). The \textsf{exoplanet} fits of the transits are overplotted (colored lines) while blue points are binned data points.}
   \label{fig:folded_transits}
\end{figure}

We used \textsf{exoplanet}, a code for probabilistic modeling of exoplanet transits \citep{DFM2021}, to measure the properties of the transiting planets. The model we set up consisted of four elements: three planet transit components with Keplerian orbits and limb-darkened transits, and a Gaussian Process (GP) component that models residual stellar variability. 
We combined the data from all the sectors into one time series with the same median normalized brightness. 
The planet models were computed with \textsf{exoplanet} using \textsf{STARRY} \citep{exoplanet:luger18}, while the GP was computed using \textsf{celerite} \citep{exoplanet:foremanmackey17,exoplanet:foremanmackey18}. 
The GP component is described as a stochastically-driven, damped harmonic oscillator with parameters of $\log(S_0)$ and $\log(\omega_0)$, where the power spectrum of the GP is
\begin{equation}
    S(\omega) = \sqrt{\frac{2}{\pi}} \frac{S_0 \, \omega_{0}^{4}}{(\omega^{2} - \omega_{0}^{2})^2 + \omega^{2} \, \omega_{0}^{2} / Q^2} \,  ,
\end{equation}
and a white noise term, with a model parameter of the log variance. We fixed Q to $1/3$ and put wide Gaussian priors on $\log(S_0)$ and $\log(\omega_0)$ with means of the log variance, and one log of one-tenth of a cycle, respectively, and a standard deviation on the priors of 10. 
This form of GP was chosen because it enables us to model a wide range of low-frequency astrophysical and instrumental signals without requiring a physical model for the observed variability. The white noise term carried the same prior as $\log(S_0)$. The planet model was parameterized in terms of limb darkening, log stellar density and stellar radius for the three planets. 
Each individual planet was parameterized in terms of log orbital period, time of transit, planet-to-star radius ratio, impact parameter, orbital eccentricity and periastron angle at the time of transit. The time of transit was set to a transit near the center of the time-series to minimize correlations between transit epoch and orbital period. The stellar radius had a Gaussian prior of $1.157 \pm 0.025 R_\odot$. The log mean stellar density, in cgs units, had a Gaussian prior with a mean of $\log{1.1}$ and standard deviation of 0.18~dex. The limb darkening followed the \citet{exoplanet:kipping13} parameterization. The log orbital periods, time of transit, and log planet-to-star radius ratio of the three candidates had Gaussian priors with means at the values listed on ExoFOP-TESS (aside from the single transiting planet where we fixed the orbital period). The impact parameter had a uniform prior between zero and one plus the planet-to-star radius ratio. Eccentricity had a beta prior with $\alpha=0.867$ and $\beta=3.03$ \citep[as suggested by ][]{Kipping2013}, and was bounded between zero and one. The periastron angle at transit was sampled in vector space to avoid the sampler seeing a discontinuity at values of $\pi$. We sampled the posterior distribution of the model parameters using the No U-turn Sampler \citep[NUTS, ][]{NUTS} which is a form of Hamiltonian Monte Carlo, as implemented in \textsf{PyMC3} \citep{exoplanet:pymc3}. We ran 4 simultaneous chains, with 3000 tuning steps, and 2000 draws in the final sample. 
The phase-folded transits of the three candidates, along with best-fitting transit models, are shown in Figure \ref{fig:folded_transits}, and the model parameters are provided in Table \ref{tab:TransitParameters}. 
Note that, by using the transit duration and depth along with stellar mass and radius, we can estimate the period through Kepler's third law in the assumption of planar and circular orbit, as shown in Section 6.5 of \citet{Seager2003}. This yields a period of approximately $46 \pm 4$ days.
A period between 43 and 46 days would also be consistent with the possible overlap of a secondary transit corresponding to momentum dump in sector 30, with a missing signal that would fall in the downlink gap of sector 31 and just short of the beginning of observations for sector 3. Moreover, if we assume that the signal at BJD at 2136.5 is a secondary transit happening at the same time of the reported momentum dump in sector 30, we find a period of 44.47 days, which is also consistent with our rough estimate. 
On the basis of \tess data only, we are not able to establish if the signal that we observed in sector 30 at BJD at 2136.5 is only due to an instrumental artifact or if it also includes the transit of planet $d$.

\subsection{Additional radial velocity data}
\label{rv_modeling}
\begin{figure}
    \centering
    \includegraphics[width=\linewidth]{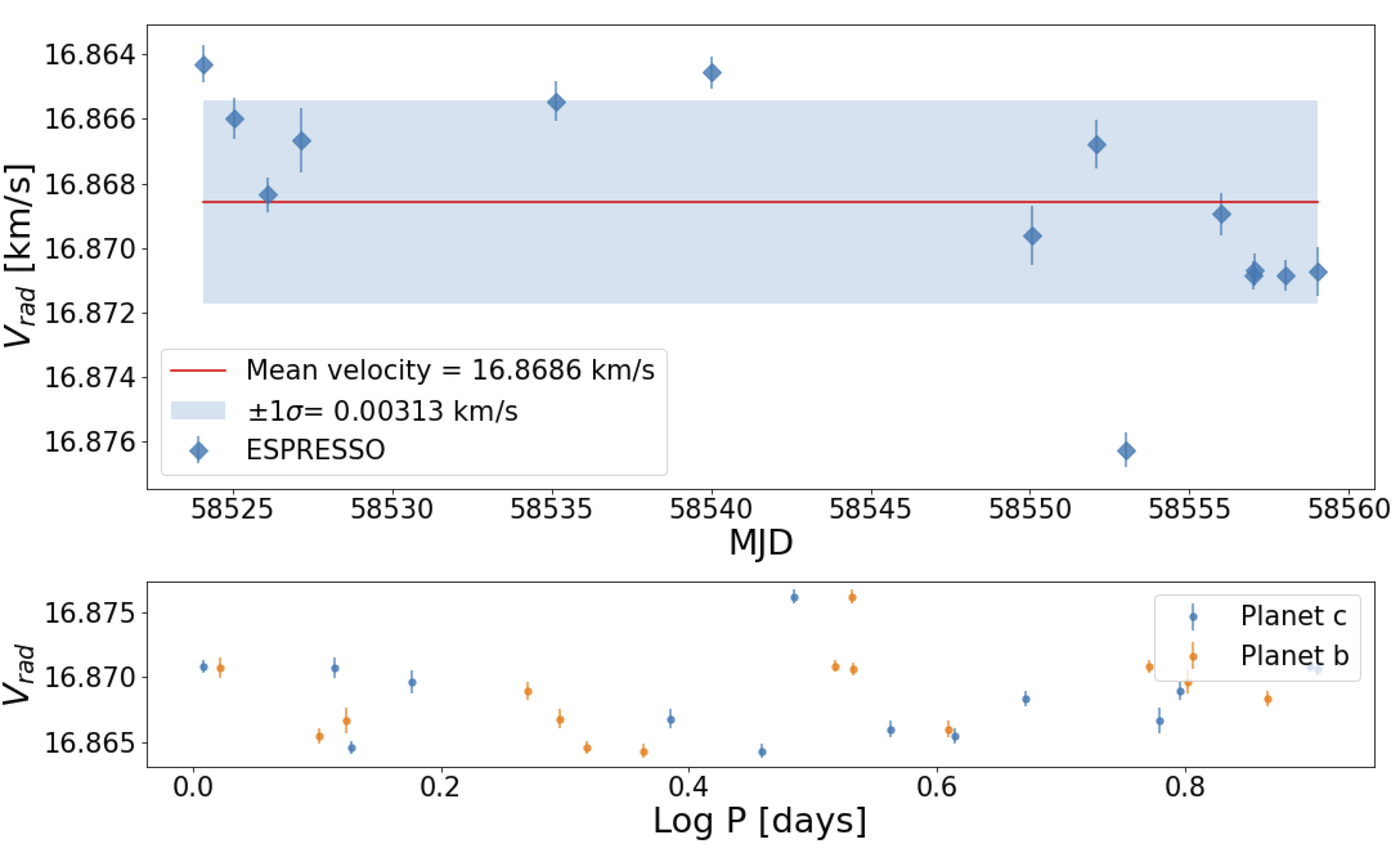}
    \caption{Upper panel: ESPRESSO precise radial velocity measurements of TOI-411, taken over February and March 2019. Lower panel: the pRVs have been phase-folded on the periods of planet candidates TOI-411 $b$ and $c$.}
    \label{fig:my_pRV}
\end{figure}

By searching the ESO archive, we found that TOI-411 was observed by the Echelle SPectrograph for Rocky Exoplanets and Stable Spectroscopic Observations (ESPRESSO; \citet{Pepe2014}) between the 10th of February and the 17th of March 2019 (program ID: 0102.C-0456; PI: Van Eylen).
ESPRESSO is a high-resolution (R$\sim$140,000) spectrograph operating at the VLT at ESO’s Paranal Observatory, capable of measuring radial velocities of bright stars (V< 8 mag) with a precision of 10 $cm/s$ in the 377–790 nm range. 
TOI-411 was observed 14 times in roughly one month, with typical exposure of 600s and SNR ranging from 120 to 243. The observations are summarised in Table\ref{tab:espresso}. We looked at the radial velocities publicly available in the archive, which are derived from the cross-correlation of the stellar spectra with a G2 template. The data is shown in Fig.\ref{fig:my_pRV}.

\begin{table}
\label{tab:espresso}
\centering
\caption{ESPRESSO radial velocity measurements of HD~22946}	
\resizebox{\columnwidth}{!}{%
\begin{tabular}{c  c  c  c}
\hline\hline
$CCF_{RV} [km/s]$ & $CCF_{RV_{err}} [km/s]$ & $<SNR>$ & $MJD$\\\\ \hline
16.864287& 0.000561& 181& 58524.05723591\\
16.865981& 0.000627& 164& 58525.05175402\\
16.868350& 0.000553& 184& 58526.09212435\\
16.866658& 0.001012& 109& 58527.12891579\\
16.865454& 0.000619& 168& 58535.12063523\\
16.864550& 0.000496& 203& 58540.03340549\\
16.869616& 0.000916& 119& 58550.06931525\\
16.866772& 0.000763& 139& 58552.06584297\\
16.876263& 0.000540& 188& 58553.01522444\\
16.868946& 0.000662& 156& 58555.99860106\\
16.870850& 0.000447& 224& 58557.00297657\\
16.870675& 0.000487& 208& 58557.06290428\\
16.870844& 0.000482& 210& 58558.02527738\\
16.870741& 0.000766& 138& 58559.03833688\\
\hline 
\end{tabular} 
}
\end{table}

We modelled the ESPRESSO radial velocity data using \textsf{exoplanet} \citep{exoplanet:foremanmackey17,exoplanet:foremanmackey18}. Specifically, we included the V$\_{\rm{rad}}$ data with the transit model and performed a joint, self-consistent model. As parts of the V$\_{\rm{rad}}$ component of the model, we included a periodic Gaussian Process with a characteristic timescale of 50 days and an amplitude of 2 m/s. The values chose were selected to be longer than any Keplerian signal from the planets, and designed to encompass any long-term variations in the data. The period of planet $d$ was fixed to 46 days.
The posterior distributions of the semi-amplitudes are consistent with non-detection of any signal in the radial velocity data, with $3\sigma$ upper limits of 4.1, 3.5, 8.4 m/s for planets $b$, $c$ and $d$, respectively. These values for the semi-amplitudes yield upper mass limits of $\sim$ 11, 14.5 and 24.5 $M_{\oplus}$. The masses are forecast to be in the region of 3.6, 8.0, and 11 $M_{\oplus}$ \citep{Chen&Kipping2017}, and therefore the non-detection of any mass signal is not unexpected.
Finally, there is evidence for a slope in the RV data, although the significance is not high. In the month that the RV data cover, the star shows show a 6 m/s change in measured radial velocity. This signal is consistent with an amplitude that might be seen if the source of the signal were to be stellar rotation or RV jitter \citep{Luhn2020}. If the source is not stellar or instrumental noise, then the slope could be due to an external body such as another planet or stellar companion. However, the available data is not sufficient to make predictions beyond saying that the data is not inconsistent with a companion more massive than Neptune on an orbital period longer than about 30 days.

\begin{table*}
\label{tab:TransitParameters}
\centering
\caption{Planet parameters. The priors for each fitted quantity are explained in the text (Section \ref{modeling}).}	
\begin{tabular}{c c c c}
\hline\hline
Parameter & Median & +1$\sigma$ & -1$\sigma$\\
\hline
&Model Parameters& \\
{\bf Star} & & & \\
${\rho}$ [g~cm$^{-3}$] & 1.10 & 0.19  & 0.16 \\
Limb darkening $u_1$ & 0.18 & 0.20 & 0.13  \\
Limb darkening $u_2$ & 0.19 & 0.28 & 0.24 \\ \\
{\bf TOI-411 b} & & & \\
${T_0}$ (BJD - 2457000) & 1385.729875 & 0.004361 & 0.001199 \\
$\ln{Period}$ [days] & 1.396319 &  0.000006 &0.000010   \\
Impact parameter & 0.29 & 0.23 & 0.12	\\
Depth (ppm) & 201 & 17& 16\\
eccentricity &  0.126 & 0.190 & 0.009 \\
$\omega$ [radians] & -0.7  & 2.2 & 2.9  \\
 & & & \\
{\bf TOI-411 c} & & & \\
${T_0}$ (BJD - 2457000)& 1386.188119 & 0.001278 & 0.001313  \\
$\ln{Period}$ [days] & 2.2589566& 0.0000027& 0.0000024 	 \\
Impact parameter & 0.42 & 0.20	&0.27	\\
Depth (ppm) & 502 & 32& 30\\
eccentricity & 0.16 & 0.17  & 0.10  \\
$\omega$ [radians] & 0.2 &  2.2 &   2.4  \\
 & & & \\
{\bf Single Transit Event} & & & \\
${T_0}$ (BJD - 2457000)& 1425.164717 & 0.002288 & 0.001901 \\
$\ln{Period}$ [days] & - &	- &-	 \\
Impact parameter & 0.35 & 0.21  & 0.23 \\
Depth (ppm) & 704 & 64& 65\\
eccentricity & -  & - & -\\
$\omega$ [radians] & - & -	& -  \\ \\
\hline\hline \\
& Derived Parameters &\\
{\bf TOI-411 b} & & & \\
Period [days] & 4.040301 & 0.000023	 & 0.000042	\\
${R_p/R_*}$ &0.01358& 0.00060	& 0.00059 \\
Radius ${[R_\oplus]}$ & 1.72 &  0.10 & 0.10	\\
${a/R_*}$ & 9.84 & 0.53 & 0.51 \\
$a$~[AU] & 0.0528& 0.0030 & 0.0030\\
Inclination (deg) & 88.3 & 1.1 & 1.2 \\
Transit duration (hours) & 2.96 & 0.38  & 0.53 \\
Equilibrium temperature (K) & 1378 & 36 & 36\\
 & & & \\
{\bf TOI-411 c} & & & \\
Period [days] & 9.573096 & 0.000026	 &	0.000023\\
${R_p/R_*}$ & 0.02161 & 0.00077& 0.00075 	\\
Radius ${[R_\oplus]}$ & 2.74 & 0.14 &  0.14 \\
${a/R_*}$ & 17.49 & 0.95 & 0.91 \\
$a$~[AU] & 0.0939 & 0.0054 & 0.0054 \\
Inclination (deg) & 88.57 & 0.86 & 0.53 \\
Transit duration (hours) & 3.95 &0.82 & 0.80 \\
Equilibrium temperature (K) & 1033 & 27 & 27 \\
 & & & \\
{\bf Single Transit Event} & & & \\
Period [days] & - & - & -	\\
${R_p/R_*}$ & 0.0255 & 0.0012	& .0012 	\\
Radius ${[R_\oplus]}$ & 3.23 &  0.19 & 0.19 	\\
${a/R_*}$ & - & -  & -\\
$a$~[AU] & - & - & - \\
Inclination (deg) & - & - & - \\
Transit duration (hours) & 6.7 & 1.0 & 1.1 \\ 
Equilibrium temperature (K) & - & - & - \\
\hline 
\end{tabular} 
\end{table*}

\section{Ruling out false positive scenarios}
\label{NEB}

TESS is designed to have a large field of view (24$^{\circ} \times $96$^{\circ}$) to survey 85\% of the entire sky within the two years of the nominal mission. 
As a consequence, it also has a large pixel scale (21" per pixel), with a focus-limited PSF that can be as large as 1'. This implies that the collected time-series photometry of a given target can be contaminated by nearby or background/foreground unresolved sources.
Specifically, the aperture used to extract the light curve could contain more sources. The light contribution of this other source could dilute the transit signal, resulting in an underestimation of the associated planetary radius. Or the additional source could be an eclipsing binary of which we can only observe one set of eclipses due to signal dilution caused by the light collected from the target star, producing a false positive. 
Therefore, the various scenarios we have to rule out include: a) possible instrumental (systematic) effects; b) the host star being in an eclipsing binary or multiple system; c) the host star being close to or aligned with a foreground or background star eclipsed by a stellar companion or transited by a planet.  

To address the first scenario, we inspected TESS Data Releases for the interested sectors to look for instrument malfunctioning and reported artifacts. 
As stated in Section \ref{sec:detection}, we found a feature at 2136.5 BTJD associated with one of the momentum dumps, i.e. times at which the reaction wheel speeds of the satellite were reset. We therefore decided to discard this event as an instrumental artifact. We did not find any other feature of similar nature in the entire light curve.

To clear HD~22946 from the remaining false positive scenarios, we first directly analyze \tess images and light curves, by using the \textsf{DAVE} pipeline \citep{Kostov2019}, as described in subsection~\ref{vetting}.

Secondly, we investigate the veracity of the transit-like signals using additional ground-based time-series and high-resolution imaging collected within TFOP, as described in subsections~\ref{highcontrast} and~\ref{lcurves}. 
All of these tests have exclusively been performed on planet candidates \textit{b} and \textit{c}, since they are based on the detection of multiple events and therefore scheduled around the predictions made from the known periods. 
However, we note that if we assume a population of false positive randomly distributed over the entire sky, a star with at least one (confirmed) transiting planet is more likely to show signals of a second transiting planet rather than being a false positive \citep{Latham2011, Lissauer2012, Guerrero2021}. This multiplicity factor increases the probability that a planet candidate is a true planet rather than a false positive by a factor of 20-50 \citep{Rowe2014,Lissauer2014,Guerrero2021}. 
These considerations help us to increase our confidence in the planetary nature of all three candidates investigated in this work.

To quantify this confidence, we computed the Bayesian probability for the signals being false positives with \textsf{triceratops} \citep{Giacalone2020} and \textsf{vespa} \citep{Morton2015} in subsection~\ref{ffp}.

\subsection{\textsf{DAVE} analysis}
\label{vetting}

In order to tackle the false-positive scenarios b) and c), we first use the \textsf{DAVE} package, following the workflow of \citet{Cacciapuoti2022}.
We use different modules of the code, with the following rationale.
\begin{itemize}
    \item \textsf{centroids} generates the in-transit and the two (before and after the event) out-of-transit images for each transit. Then, it subtracts the overall in-transit image from the overall out-of-transit image to produce a difference image. Finally, it measures the center-of-light for each difference image by fitting the \tess Pixel Response Function (PRF\footnote{For more information about \tess PRF see \url{https://archive.stsci.edu/files/live/sites/mast/files/home/missions-and-data/active-missions/tess/_documents/TESS_Instrument_Handbook_v0.1.pdf}})to the image and computes the overall photocenter by averaging over all the events examined at the previous step. 
    The aim of this procedure is to pinpoint the true source of the transit-like events. See \citet{Kostov2019} for further details.
    
    \item \textsf{modelshift} phase-folds the light curve and convolves it with the best-fit trapezoid transit model, thus highlighting light curve features. Furthermore, it shows the average of the input primary signals, the average of the odd and even signals, the most prominent secondary, tertiary and positive features. These results allow us to investigate the possibility that the source of the signal is an eclipsing binary. See \citet{Kostov2019} for further details.
    
   \item \textsf{DAVE} finally runs the \textsf{astropy}-implemented Lomb-Scargle periodogram and generates a PDF file showing both the light curve phase-folded on the period of the inspected signal and the LS period. This test is performed to check whether light curve modulations occur on (half) the orbital period of the candidate. This is typical of beaming, reflection and ellipsoidal effects of binary systems \citep[see e.g.,][]{Morris1993, faigler2011,Shporer2017}.
\end{itemize}

\textsf{DAVE} results for both TOI~411.01 and TOI~411.02 show no significant additional eclipse in the light curves, no odd-even differences in consecutive signals, nor flag the shapes of the transits. Even though some transits of planet $b$ and $c$ produce unreliable photocenter difference images due to their low signal-to-noise ratio, the ones that are reliable confirm that the target star is the true source of the signal. 
For every sector, the average of all photocenters for each difference image of planets $b$ and $c$ falls on the target star. Since no significant shift is found, HD~22946 can be considered the source of the recovered signals, at TESS resolution limit.
Regarding planet candidate $d$, no reliable photocenter measurement can be drawn due to the fact that only a single event has been observed by \tess. \textsf{Modelshift} also requires periodic signals, so it cannot be applied to the single-event candidate $d$.

Several resulting images of the \textsf{centroids} module and examples of \textsf{Modelshift}'s results are shown in the Appendix (Fig. \ref{fig:centroidsb}, Fig. \ref{fig:centroidsc}, Fig.\ref{fig:modshiftb} and Fig.\ref{fig:modshiftc}).

Finally, it is worth mentioning that TOI-411~$b$ passed all the automated SPOC pipeline Data Validation (DV) diagnostic tests but the difference-image centroiding test for sectors 4 and 31. However, the difference-image centroiding tests for this planet are affected by the fact that TOI-411 appears to be slightly saturated and the SNR of the transit signal is low ($\sim10$). TOI-411~$c$ passed all the DV diagnostic tests, including the difference-image centroiding tests (but sector 4). Both transit signatures passed the odd/event transit depth tests reported by the DV. The summary of these results can be found on ExoFOP-TESS\footnote{https://exofop.ipac.caltech.edu/tess/target.php?id=100990000}.

\subsection{Ruling out nearby eclipsing binaries}
\label{lcurves}

\begin{figure}
\centering
\includegraphics[width=\linewidth]{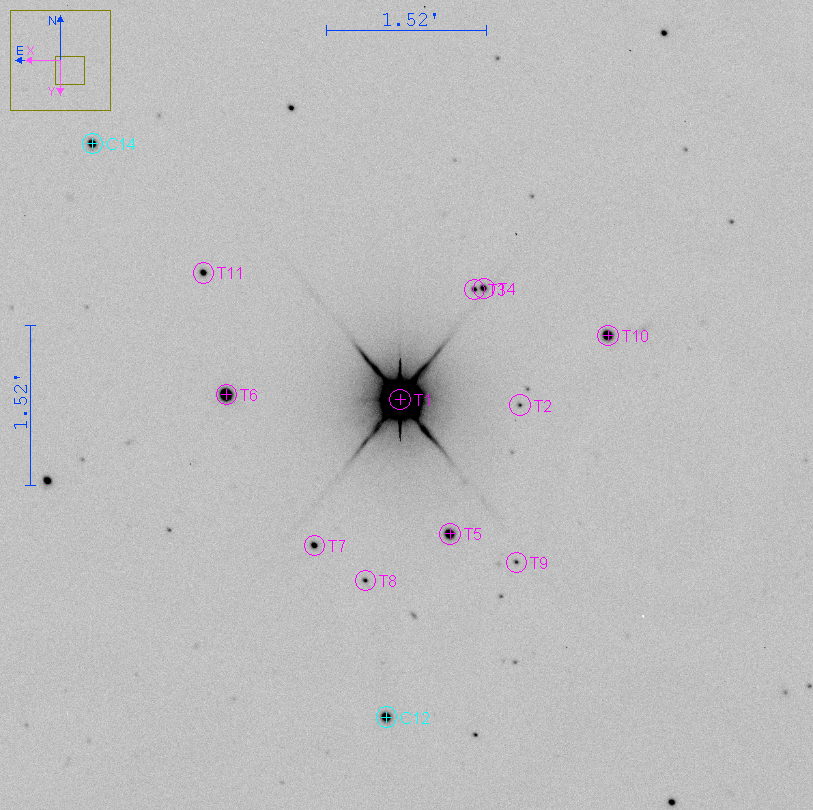}
\caption{Las Cumbres Observatory image of the host star HD~22946 in the $rp$ filter. The 10 sources identified by pink circles as T2 through T11 are within a 2.5' radius of TOI~411 and are potentially bright enough (see text) to cause the TOI~411 $b$/$c$ signals. The nine closest nearby sources are also displayed in Fig.\ref{fig:tpf} and fall just outside the automatic, SPOC-based aperture used in Sector 4.}
\label{fig:neb}
\end{figure}

We acquired ground-based time-series follow-up photometry of the stars in the field around TOI-411 as part of the \tess Follow-up Observing Program \citep[TFOP;][]{collins:2019}\footnote{https://tess.mit.edu/followup}. Observations were scheduled to cover the times of transit of TOI-411.01 and TOI-411.02, predicted by the initial published Quick Look Pipeline (QLP, \citep{Huang2020}) ephemerides from \tess sectors 3 and 4 using the {\tt TESS Transit Finder}, which is a customized version of the {\tt Tapir} software package \citep{Jensen:2013}. If the events detected in the \tess data are indeed on-target, the shallow QLP reported depths of 280 ppm and 166 ppm for TOI-411.01 and TOI-411.02, respectively, would not generally be detectable in ground-based observations. Instead, we saturated the bright star TOI-411 to enable the extraction of light curves of nearby fainter stars to attempt to rule out or identify nearby eclipsing binaries (NEBs) as potential sources of  \tess detection.

We observed an ingress plus about 50\% of a predicted TOI-411.01 event and a full transit of TOI-411.02 using the Las Cumbres Observatory Global Telescope \citep[LCOGT;][]{Brown:2013} 1.0\,m network nodes at the South Africa Astronomical Observatory and Cerro Tololo Inter-American Observatory, respectively. The TOI-411.01 observation was on UTC 2019 February 24 in Sloan $r'$ band and the TOI-411.02 observation was on UTC 2020 December 02 in Sloan $i'$ band. The 1\,m telescopes are equipped with $4096\times4096$ SINISTRO cameras having an image scale of $0\farcs389$ per pixel, resulting in a $26\arcmin\times26\arcmin$ field of view. The images were calibrated by the standard LCOGT {\tt BANZAI} pipeline \citep{McCully:2018}. Photometric data were extracted using {\tt AstroImageJ} \citep{Collins2017}. 

Additionally, full transit duration observations were performed with the 0.305m Perth Exoplanet Survey Telescope (PEST) in the $Rc$ filter on the nights of 2019-02-16 (TOI-411.02) and 2020-12-06 (TOI-411.01). The data reduction and the aperture photometry were performed by using a custom pipeline based on C-Munipack\footnote{http://c-munipack.sourceforge.net/}.

Although the original analyses were based on the initial published QLP ephemerides from \tess sectors 3 and 4, we reanalyzed the data relative to the TOI-411.01 and TOI-411.02 ephemerides derived in this work. We checked for possible NEBs that could be contaminating the SPOC and QLP photometric apertures, which generally extend $\sim1\arcmin$ from the target star. To account for possible contamination from the wings of neighboring star PSFs, we searched for NEBs in all known Gaia EDR3 and TIC version 8 nearby stars out to $2\farcm5$ from TOI-411 that are possibly bright enough in \textit{TESS} band to produce the \textit{TESS} detection (assuming a 100\% eclipse and 100\% contamination of the \textit{TESS} aperture). In order to account for possible delta-magnitude differences between \textit{TESS}-band and the follow-up filter bands, we checked stars that are an extra 0.5 magnitudes fainter in \textit{TESS}-band than needed. We consider a star cleared of an NEB if the RMS of its 10-minute binned light curve is more than a factor of 5 smaller than the adjusted expected NEB depth in the star (adjusted to allow for the potential \textit{TESS}-band delta-magnitude difference). We then visually inspect each neighboring star's light curve to ensure no obvious eclipse-like signal.

The transit depths we derive in this work are deeper than the initial QLP depths. To be conservative when checking for potential NEBs, we used the shallower TOI-411.02 QLP depth of 166 ppm to calculate adjusted expected depths in all of the nearby stars. This resulted in a check of the 10 stars labeled T2 through T11 in Figure \ref{fig:neb}. To simplify the presentation of the results, we also checked the same 10 stars for the deeper transit of TOI-411.01. The faintest star we checked is T2 (TIC 100990001), which has a delta-\textit{TESS}-band magnitude of 9.86 according to TIC version 8. Assuming its true delta magnitude is 9.36 in the follow-up filter bands, a 92\% eclipse in T2 would produce a 166 ppm depth in TOI-411. We find that the RMS of all 10 light curves for each planet candidate are more than a factor of 5 smaller than the adjusted expected NEB depth in the respective star. All of our follow-up light curves and supporting results are available on the {\tt EXOFOP-TESS} website\footnote{https://exofop.ipac.caltech.edu/tess}.


Finally, we note that no observations could be carried out to perform a similar analysis for planet candidate $d$, since the orbital period is unknown and hence we cannot make reliable predictions for the epoch of the next transits.


\subsection{Ruling out unresolved stellar companions}
\label{highcontrast}
\begin{figure}
    \centering
    \includegraphics[width=\linewidth]{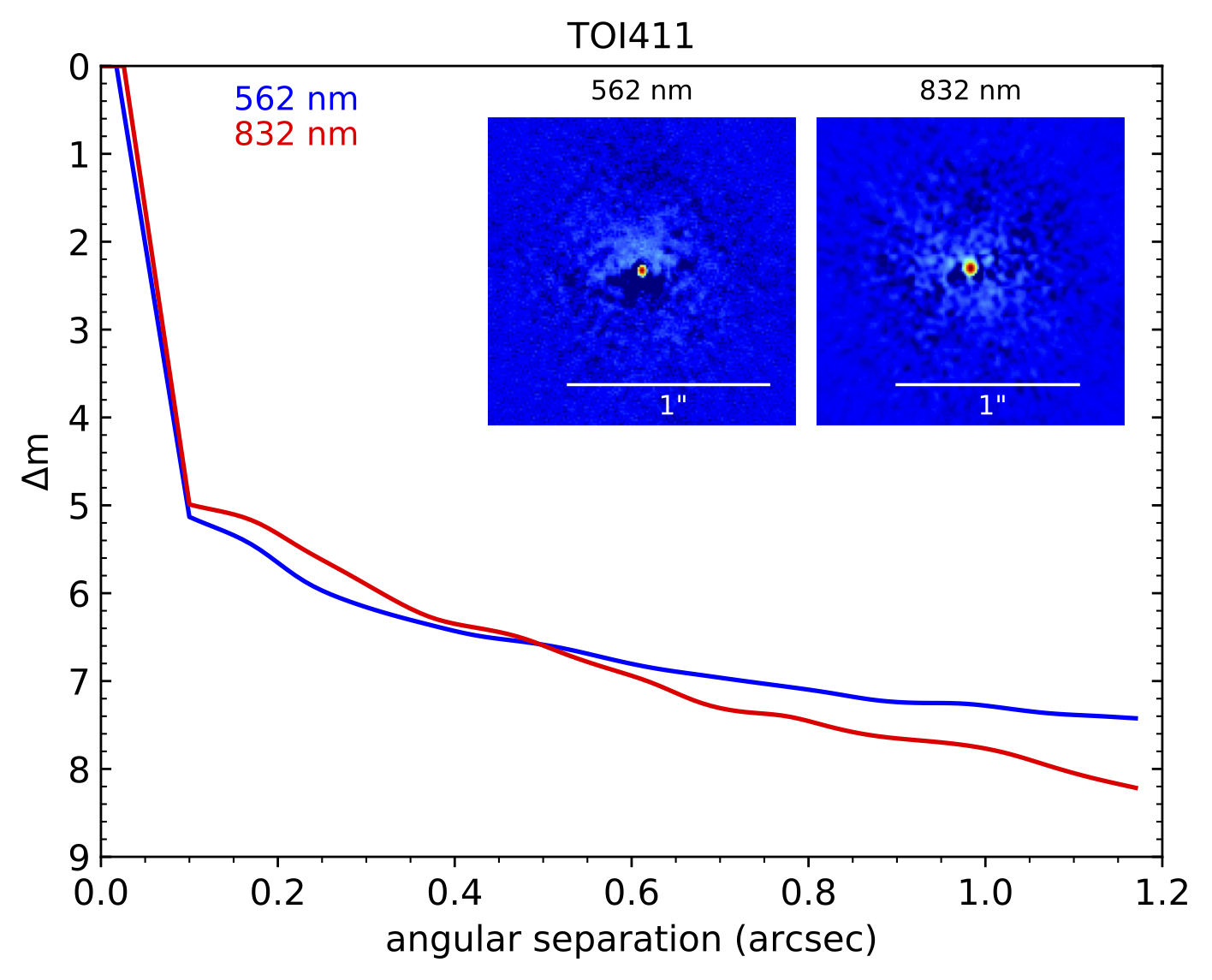}
    \caption{Gemini-South Zorro's speckle high-contrast imaging observations of TOI-411 on UT 2020-11-25: 5$\sigma$ sensitivity curves and reconstructed images. 
    The data show that there are no close-in companions detected down to 5 magnitude fainter than our target at a separation of 0.1 arcseconds.}
    \label{fig:gemini}
\end{figure}

\begin{figure}
    \centering
    \includegraphics[width=\linewidth]{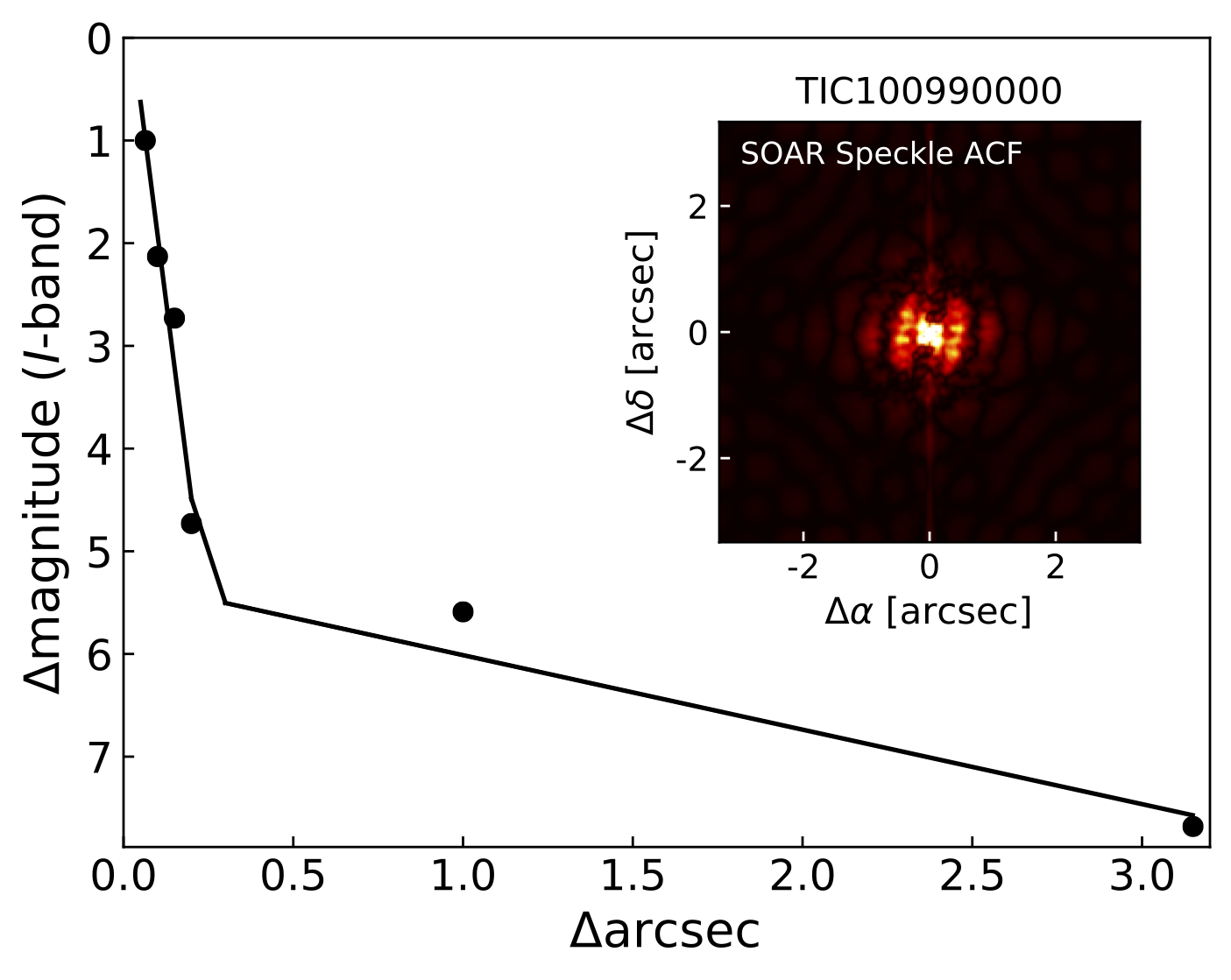}
    \caption{Analysis of the SOARCam high contrast image of TOI~411:
    the two-dimensional autocorrelation function and the reconstructed image of the field are shown. 
    The data show that there are no close-in companions detected within 3 arcseconds of TOI~411, which would show up as additional peaks in the autocorrelation function.}
    \label{fig:soar}
\end{figure}

Unresolved, dimmer stars could fall in the same 21''-wide TESS pixel. If an additional source is indeed present, and if this source is an eclipsing binary, its (diluted) eclipses might be the signals we observe in the light curve. Besides, the light contamination due to this unresolved star (system) might dilute the transit candidates, resulting in underestimated radii for the exoplanets \citep[see e.g.,][]{Furlan2017,Ciardi2015}.
To rule out stellar companions and foreground/background stars at close separations unresolved in \tess images, we obtained speckle imaging observations with the 8m Gemini-South telescope's Zorro instrument \citep{Scott2021}, which provides speckle imaging simultaneously in two bands centered at 562~nm and 832~nm, resulting in reconstructed images and robust contrast limits on companion detections \citep[see][]{Howell2011,Howell2016}. TOI~411 was also observed with the 4.1m SOuthern Astrophysical Research (SOAR) telescope \citep{Ziegler2020, Ziegler2021} speckle imaging camera HRCam \citep{Tokovinin2018}.
Specifically, HD~22946 was observed with Zorro on UT 2020-11-21 and 2020-11-25 in both bands. The estimated PSF at the time of observations was 0.02". The 5$\sigma$ sensitivity curves and the reconstructed images for the two bands are shown in Figure \ref{fig:gemini}. We find that TOI-411 has no companion brighter than about 5 to 8 magnitudes, respectively, from about 0.1" out to 1.2".
Observations with the HRCam on the SOAR were carried out on the UT 2019-02-18 in the I-band, centered at 879nm. The estimated PSF at the epoch of the observations was 0.06". The auto-correlation functions for the data obtained with SOAR observations is shown in Fig.\ref{fig:soar}. No nearby star is detected within 3" of TOI-411, which corresponds to 188 AU at the distance of the target.
Finally, we note that the host star HD~22946 was also analyzed by \citet{Kervella2019,kervella2022} for proper motion anomalies that could be induced by an unresolved orbiting companion, resulting in a non-detection. 
The Gaia Renormalised Unit Weight Error (RUWE) of the star is indeed 1.039, a typical value of single bright stars \citep{gaia_edr3,Ziegler2020}, thus further confirming the single-star scenario.


\begin{figure*}
    \centering
    \includegraphics[width=0.9\textwidth]{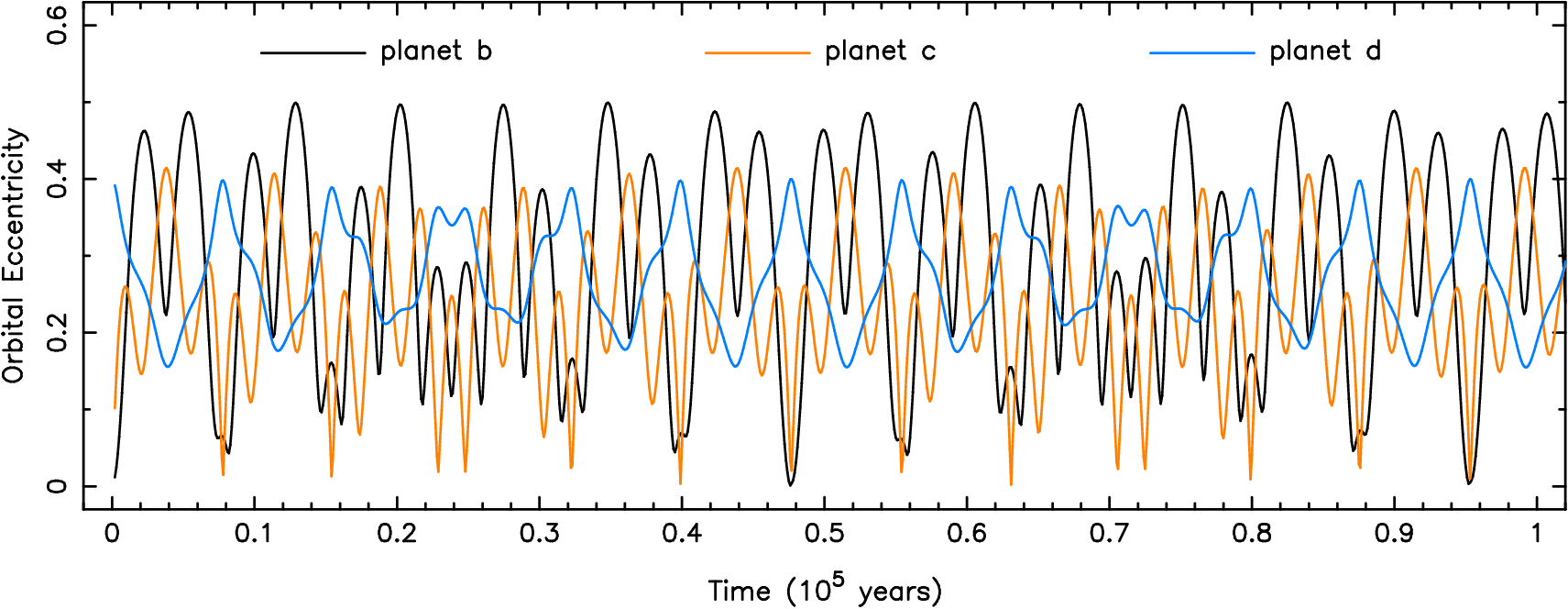}
    \caption{One of the simulated eccentricity evolution of the HD~22946 system over $10^5$ years. In the case shown here, we assumed the following extreme initial conditions for planet $d$: initial semi-major axis of 0.19~AU and eccentricity of 0.4.}
    \label{fig:dynamics}
\end{figure*}

\subsection{Quantifying the false positive probability for different scenarios}
\label{ffp}

To quantify the false positive probability  (FPP) for this system, we use the \textsf{triceratops} \citep{Giacalone2021, Giacalone2020} and \textsf{vespa} \citep{Morton2015} packages. \textsf{triceratops} is an algorithm that rules out astrophysical false positives by calculating and comparing the probabilities of various transit-producing scenarios. 
\textsf{triceratops} is specifically designed for \tess observations that consider transit scenarios originating from the target star, sources unresolved with the target star, and known nearby stars within 2.5 arcminutes from the target star. This tool encapsulates the total probability that a planet candidate is a false positive in the false positive probability (FPP) and the probability that the planet candidate is a false positive originating from a known nearby star in the nearby false positive probability (NFPP). For a planet candidate to be validated, it must achieve ${\rm FPP} < 0.01$ and ${\rm NFPP} < 0.001$ \citep[see][for more details]{Giacalone2021}. As an additional constraint in our calculations, we fold in the speckle imaging follow-up observations discussed in the previous Section. Because these observations reveal no previously unresolved companions within their detection limits, incorporating the follow-up reduces the calculated probability of the transit originating from a bound or chance-aligned star within the resolution limits of the target star, thereby reducing the FPP of the planet candidate. 
We note that, since the true period of planet $d$ is unknown, we cannot perform this kind of analysis for it.

Otherwise, we ran \textsf{triceratops} for both planet candidates $b$ and $c$ twenty times and calculated the mean and standard deviation of the resulting FPPs and NFPPs. We found ${\rm FPP} = (0.88 \pm 1.28) \times 10^{-7}$ and ${\rm FPP} = (3.56 \pm 1.27) \times 10^{-4}$ for TOI-411.01 and TOI-411.02, respectively. Because \texttt{triceratops} determines that no nearby stars are capable of being sources of astrophysical false positives, we find ${\rm NFPP} = 0$ for both candidates. Based on these results, we consider planets $b$ and $c$ to be validated.

Additionally, we used \textsf{vespa} to independently validate the same signals. \textsf{vespa} compares transit signals to a number of false-positive scenarios including an unblended eclipsing binary (EB), a blended background EB, a hierarchical companion EB, and the ‘double-period’ EB scenario. 
It accounts for the period, depth, duration, and shape of each signal as well as the colors of the target star, spectroscopic and imaging follow-up observations, and simulations of the population and distribution of field stars and binary stars at the target's position. 
We ran \textsf{vespa} on the \tess light curves to calculate the FPP for each individual planetary signal after masking additional potential transits for each target signal. The observational constraints imposed by Gemini-South Zorro's high contrast imaging of TOI~411 were included in this case as well. \textsf{vespa} is particularly sensitive to the constraint on the maximum depth of potential secondary eclipses and the maximum radius at which \textsf{vespa} considers the influence of background binary systems. 
We used the \textsf{DAVE} analysis to estimate the secondary depth constraint and set the maximum radius parameter at 21" to emulate the size of a \tess pixel. Using these inputs, we calculated an FPP of 7.77$\times10^{-12}$ and 4.51$\times 10^{-5}$ for TOI~411.01 and TOI~411.02, respectively. 
These values are well below the commonly suggested 0.01 threshold required to statistically validate these candidates as planets, in agreement with the results from \textsf{triceratops}.

\section{The HD~22946 planetary system}
\label{syst}
In this section, we present the main properties of the planetary system orbiting HD~22946. The three planets are a $1.72 \pm 0.10$ $R_\oplus$ Super-Earth (planet $b$), a $2.74 \pm 0.14$  $R_\oplus$ (planet $c$) and a $3.23 \pm 0.19$ $R_\oplus$ (planet $d$) sub-Neptunes. In subsection \ref{masses}, we estimate the planet masses and the expected semi-amplitudes of radial velocity curves that could be measured.

In subsection \ref{dynamics}, we perform N-body simulations to assess the dynamical stability of the system, while in subsection \ref{undetected} we investigate the dynamical packing of the system to determine whether additional planets might be orbiting HD~22956. 
Finally, we evaluate the capability of spectroscopic follow-ups with the JWST NIRISS instrument in subsection \ref{atmo} for the detection of possible atmospheres.

\subsection{Planet mass estimates}
\label{masses}

While TOI-411 has been observed with ESPRESSO and a total of 14 radial velocity measurements are available as explained in Section 3.4, we note that a significantly larger number of spectra would be needed to place meaningful mass constraints on the planets of the TOI-411 system.
Given the lack of them, we use empirical radius-mass relations to estimate the planets' masses.

We make use of two different equations for the inner planet and the outer ones. In fact, while the well-known \citet{Chen&Kipping2017} relations have been widely used in the literature to estimate the masses of exoplanets of any size, \citet{Otegi2020} showed that a slightly different equations better constrains Super-Earths ($1.5 < R/R_{\oplus} < 2$). Thus, adopting the planet radii in Table\ref{tab:TransitParameters} and using \citet{Chen&Kipping2017} relations, we estimate the masses of planets $c$ and $d$ to be $M_c = 7.96 \pm 0.69 M_{\oplus}$ and $M_d = 10.53 \pm 1.05 M_{\oplus}$, respectively. On the other hand, considering the corrections of \citet{Otegi2020}, we estimate a mass of $M_b = 6.29 \pm 1.30 M_{\oplus}$ for planet $b$.

Given the brightness of the host star ($V \approx$ 8.3 mag), these planets are optimal targets for mass measurements by means of radial-velocity curves.
In fact, the computed mean masses yield $V_{\rm{rad}}$ semi-amplitudes of $2.36 \pm 0.48$ m/s, $2.24 \pm 0.19$ m/s and $1.76 \pm 0.18$ m/s, respectively, which
are well within the capabilities of current Southern Hemisphere instruments such as HARPS \citep{Pepe2002} and ESPRESSO \citep{Pepe2014}, assuming an intensive monitoring.

\subsection{Transit timing variation analysis}

In order to obtain independent estimates of the planetary masses, we performed a transit timing variation (TTVs) analysis, which is also able to find possible undetected planets.
TOI-411 system was modeled with the \textsf{TRANSITFIT5} transit modeling software \citep{Rowe:2015,Rowe:2016}.
We computed a multi-planet fit that includes all 3 planets and assumes non-interacting Keplerian orbits. The best-fit model was then used to create a theoretical light curve with 2 of the other planets removed, thus isolating a single planet. The single-planet light curve was then used to measure the center of transit times for each transit event. We did not find any obvious sign of TTVs.
The period ratio of planets $b$ and $c$ are not close to strong orbital resonance and the potentially long period of $d$ suggests all three planets are not strongly dynamically coupled.  



\subsection{Dynamical stability}
\label{dynamics}

In order to validate the derived planetary architecture of the system, we performed a set of dynamical simulations to investigate the dynamical integrity of the orbits and possible eccentricity variations induced by the proximity of the planets to each other. 

We performed one simulation including only planets $b$ and $c$, and a set of different simulations on a range of possible values for the semi-major axes of planet $d$, given its uncertain location. We adopted the stellar parameters shown in Table~\ref{tab:stellarparameters} and the planetary properties provided by Table~\ref{tab:TransitParameters}.

All simulations were carried out using the N-body integration capabilities of the Mercury Integrator Package \citep{chambers1999} and adopting a methodology similar to the one described by \citet{kane2014b,kane2016d}. 
They were carried out on a hybrid symplectic/Bulirsch-Stoer integrator with a Jacobi coordinate system \citep{wisdom1991,wisdom2006b}, with a time resolution of 0.2~days to ensure adequate sampling of the planetary orbits, on a time span of $10^7$ years.

Results from the simulation of planets $b$ and $c$ alone show that they are in exceptionally stable orbits that remain approximately circular, inferring minimal dynamical interactions between them. 

To study the effects of planet $d$, we assumed a value for the semi-major axes in the range from 0.19 to 0.26 AU, corresponding to the mean and $-1\sigma$ values for the constrained period of $46 \pm 4$ days. We only considered the $-1\sigma$ (and not the $+1\sigma$) scenario to put ourselves in the most dynamically extreme situation for which the planet orbits as close as possible (within our predictions) to the other two celestial bodies and its gravitational influence has a larger effect.
The system remained stable through the semi-major axis range for planet $d$ and for the full duration of the simulations. 
We further investigated the effects of high eccentricities for planet $d$ and found that eccentricities as high as 0.4 still allow for fully stable orbits within the system even in the most compact configuration.

For example, Figure~\ref{fig:dynamics} shows the eccentricity evolution of the three planets for the extreme scenario where planet $d$ is located in a 0.19~AU orbit with a starting eccentricity of 0.4. Remarkably, the stability of the system enables a seamless transfer of angular momentum between the planets in long-term sustainable cycles. 
These results show that our understanding of the HD~22946 planetary system is dynamically sustainable, strengthening the self-consistency of our analysis.

\begin{figure}
    \centering
    \includegraphics[width=\linewidth]{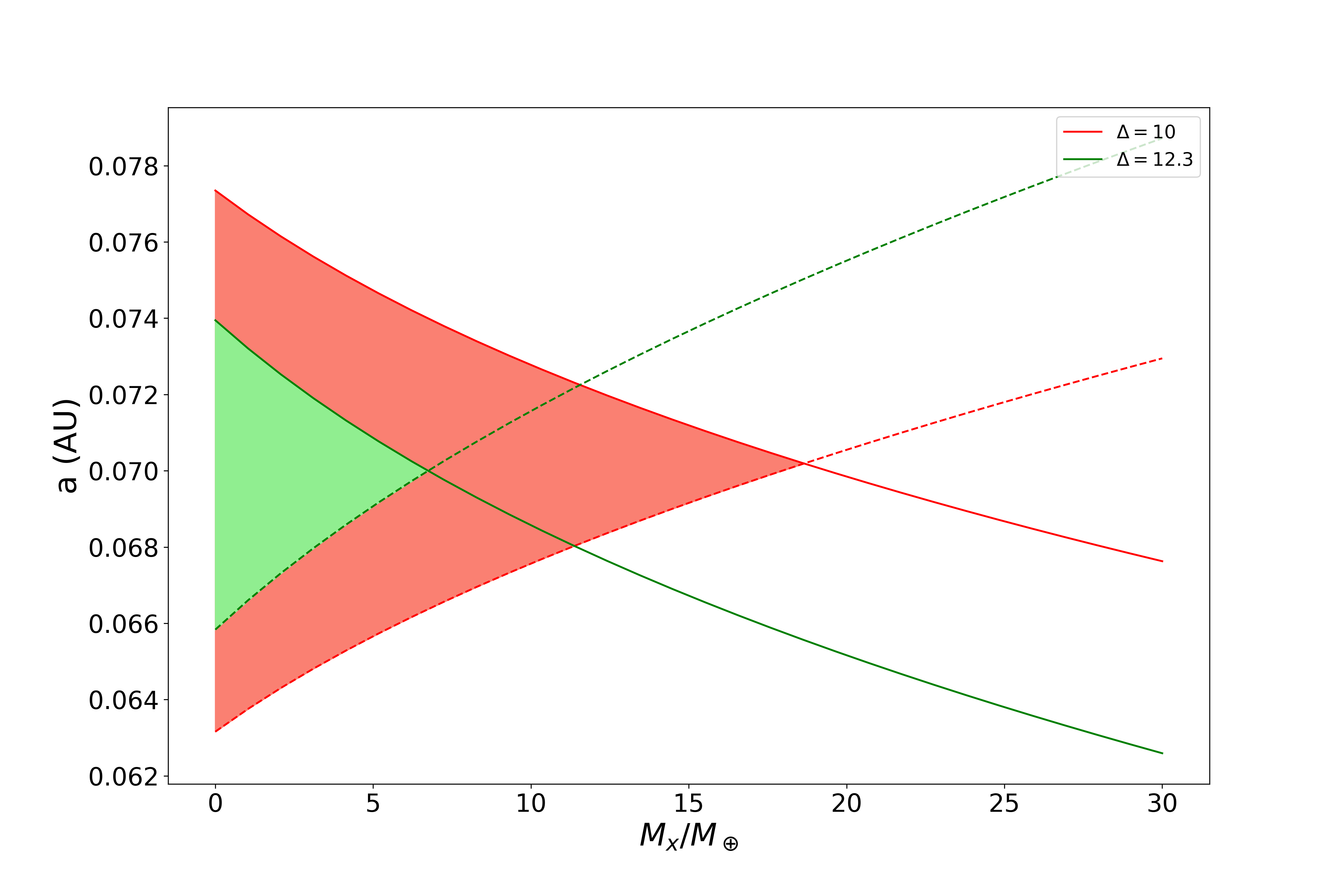}
    \includegraphics[width=\linewidth]{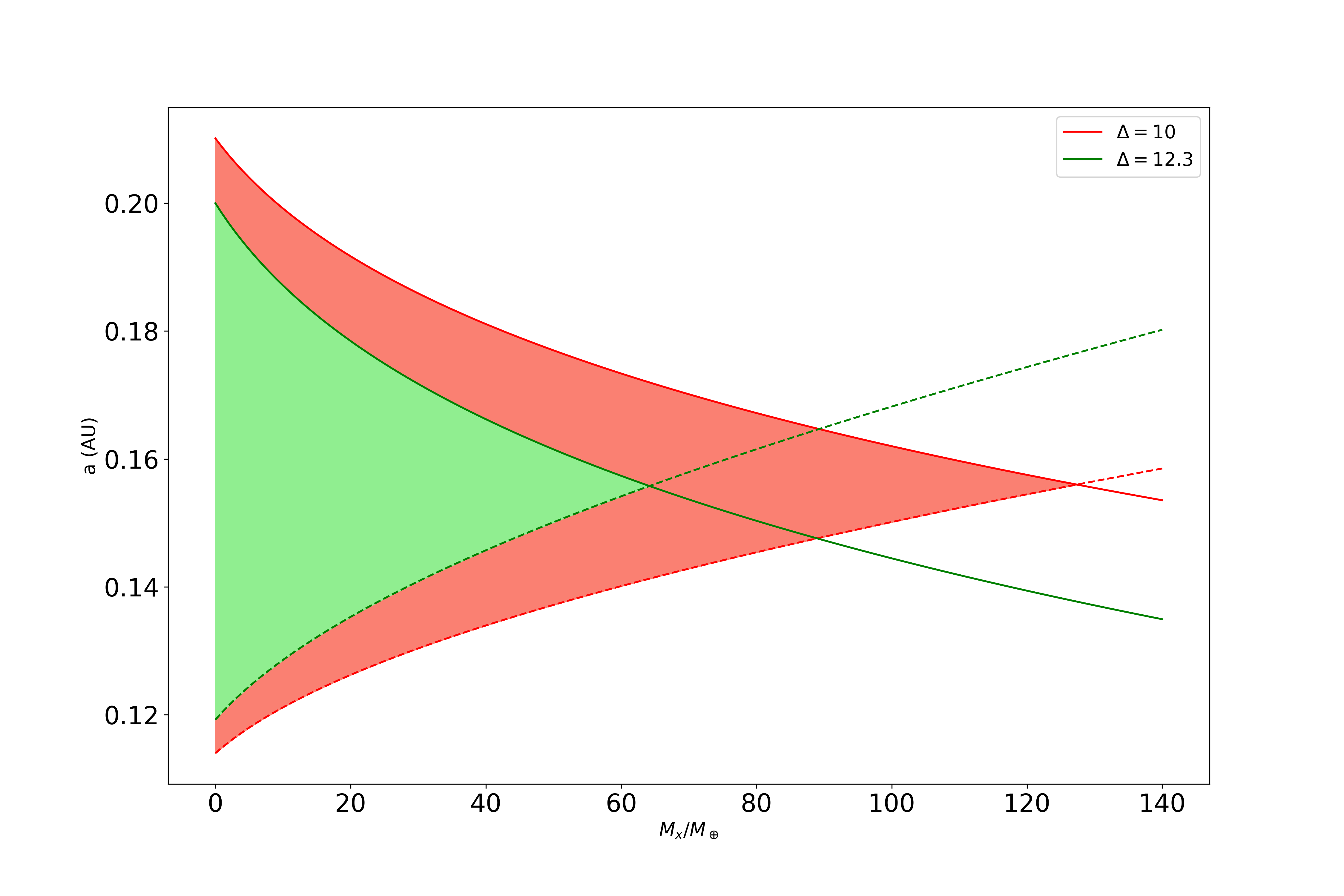}
    \caption{Mass and semi-major axis of a possible undetected planet between TOI-411 $b$/$c$ (upper panel) and TOI-411$c$/$d$ (lower panel). Minimum semi-major axis ($a_{xmin}$) in dashed line and maximum one ($a_{xmax}$) in continuous line for $\Delta_{crit}=10$ and $12.3$ in red and green respectively. Given $\Delta_{crit}$, the shaded region contains the possible combinations $(M_{x},a_{x})$ for a stable system configuration hosting one more planet.}
    \label{fig:gap}
\end{figure}

\subsection{Possible undetected planets}
\label{undetected}

Here, we investigate the presence of potentially undetected planets orbiting in the gaps between the planets presented in this work following the same approach of \citet{humphrey2020predicting}. 
\citet{Gladman1993} defined a system of two planets as Hill-stable (unstable) if $\Delta$, i.e. the separation between the two, normalized by the mutual Hill-radius ($R_H$) of the pair, is above (below) an analytically defined critical level, $\Delta_{crit}$. 
In the more general case of multi-planetary systems, \citet[]{chambers1996stability, pu2015spacing} used numerical simulations to retrieve the minimum value of $\Delta_{crit}$ above which the pair can be considered dynamically stable. 
\citet{chambers1996stability} simulated systems with circular, coplanar orbits, and found that a $\Delta_{crit} =$ 10 can be generally used to measure the stability of multi-planet systems within the described prescriptions. 
\citet{pu2015spacing} found the same $\Delta_{crit}$ as \cite{chambers1996stability} for systems on circular and coplanar orbits but extended their simulations by including systems eccentricities and inclinations drawn from a Rayleigh distribution. 
They find that planet pairs with $\Delta < \Delta_{crit} =$ 12.3 are expected to experience close encounters, orbit crossing events and/or ejections due to strong gravitational interaction between the planets.
Considering these two values for $\Delta_{crit}$, we can check if the HD~22946 system could host, between each planet pair, an additional planet without jeopardizing the system stability.
Following \citet{humphrey2020predicting}, we computed the interval ([$a_{xmin},a_{xmax}$]) of possible semi-major axes for a potential additional planet in a range of possible masses $M_x$ for both the $b/c$ and the $c/d$ pair. If, for a given mass, $a_{xmin}\leq a_{xmax}$, the planet pair is considered $unpacked$, i.e. there is enough space to host an additional planet that would not destabilize the system.
By solving equations (2) and (3) of \citet{humphrey2020predicting}, one can compute ($M_{xmax}$, $a_{x}$), the maximum mass that a potential undetected planet could have without destabilizing the system and its related semi-major axis. 
We computed $a_{xmin}$ and $a_{xmax}$ for each value of $\Delta_{crit}$ for both couples in the system. 
The results are shown in Figure \ref{fig:gap}. Both planet pairs are unpacked for each value of $\Delta_{crit}$, though there is quite limited space for an additional planet between TOI-411 $b$ and $c$. The TOI-411 $c$ and $d$, on the other hand, might host an additional planet as large as 0.4 $M_J$ based on the least stringent $\Delta_{crit}$ value of 10.
Table \ref{tab:gap} shows the computed values for $M_{xmax}$ and related $a_{x}$ under the two different $\Delta_{crit}$ conditions. 
We searched the light curve with the TLS pipeline in the intermediate range of periods and found no evidence for additional transiting candidates. However,  
there is the possibility that a potentially undetected planet is not observed transiting the star due to its orbital inclination and might be discovered via precise radial velocity techniques as in, e.g., \citet{Demangeon2021}. 
\begin{table}
\centering
\begin{tabular}{lll}
\hline
Pair b/c & $M_{xmax}/M_{\oplus}$ & $a_{x} (AU)$ \\ \hline
$\Delta_{crit}=10$ & 18.6 & 0.0698 \\
$\Delta_{crit}=12.3$ & 6.7 & 0.0693   \\
\hline
Pair c/d & $M_{xmax}/M_{\oplus}$ & $a_{x} (AU)$ \\ \hline
$\Delta_{crit}=10$ & 127.4 & 0.156 \\
$\Delta_{crit}=12.3$ & 64.2 & 0.156   \\
\end{tabular}
\caption{Maximum mass $M_{xmax}$ and related semi-major axis $a_{x}$ of a potential additional planet orbiting between planets $b/c$ or $c/d$. Two different values for $\Delta_{crit}$ are considered (see text).}
\label{tab:gap}
\end{table}


\subsection{Potential for atmospheric characterization}
\label{atmo}

\begin{figure}
    \centering
    \includegraphics[width=\linewidth]{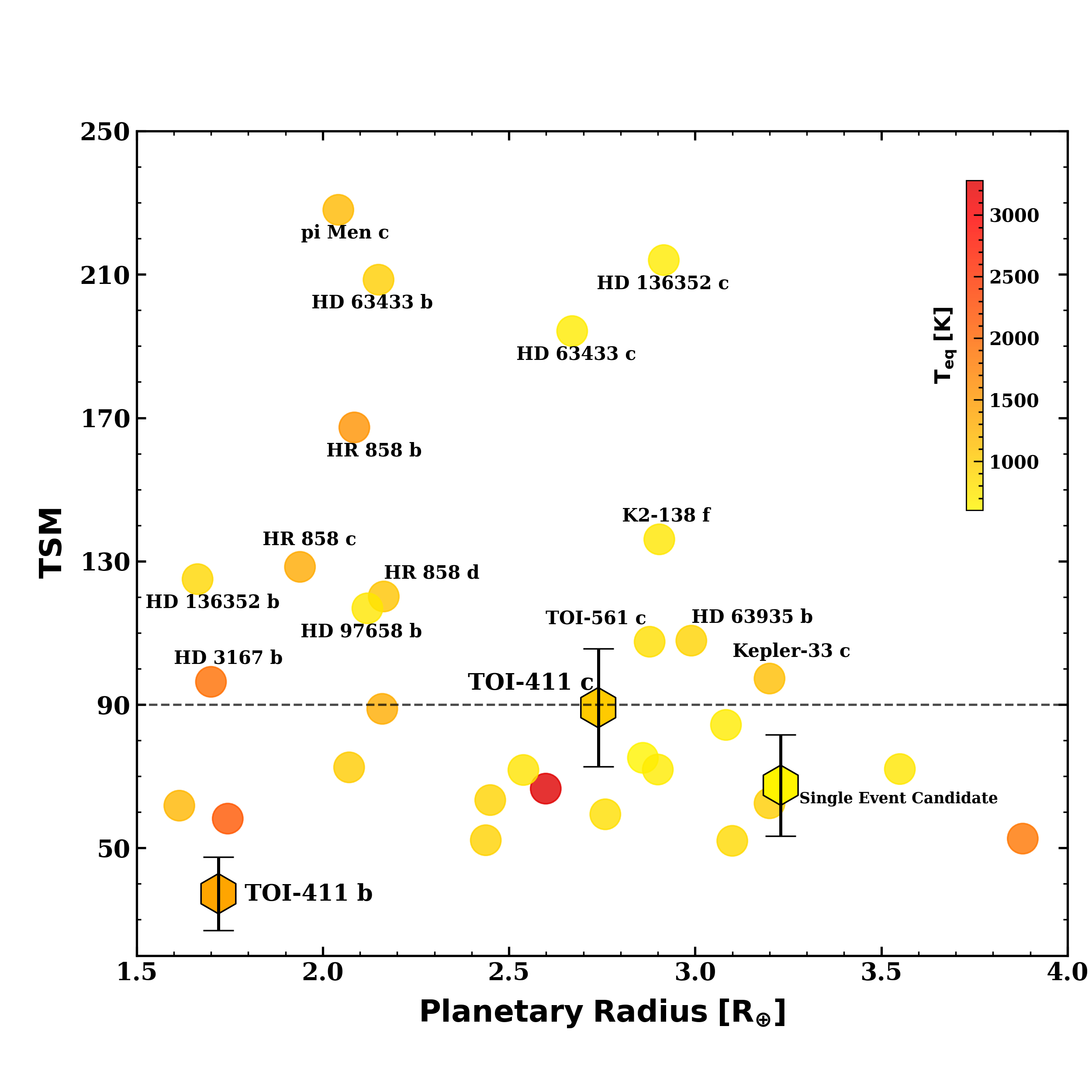}
    \caption{Transmission Spectroscopy Metric (TSM) values for known exoplanets orbiting F/G stars in less than 50 days. We only show exoplanets for which TSM$>50$ for clarity as $\sim$ 98\% of worlds in the sample lays below this value. While TOI-411 $b$ and $c$ fall in the 2\% of this kind of planet with TSM$>50$, only TOI-411 $c$ passes the suggested threshold of 90, joining an elite group of 15 super-Earths and sub-Neptunes orbiting Sun-like stars amenable for spectroscopic follow-ups with JWST.}
    \label{fig:TSM}
\end{figure}

The observed period and derived radius for TOI-411 $b$ place it in the hot super-Earths range. Specifically, in the $1.5-2.0 R_{\oplus }$ interval where a scarcity of exoplanets has been observed by \citet{Fulton2017}. This is considered to be a transition region from rocky worlds, with a high-molecular-weight atmosphere, to low-density worlds, dominated by a H/He gaseous envelope. The former may have accreted a primary atmosphere in the early stages of formation that might have been dispersed due to escape processes, such as stellar radiation photo-evaporation \citep{Fulton2018} or core-powered mass loss \citep{Ginzburg2018}. The only super-Earth with a detected (and largely discussed) atmosphere is 55 Cnc $e$, which probably lost its primary H/He envelope \citep{zhang2021no}. Even if TOI-411 $b$ might have lost its primary atmosphere during its early evolution, a large reserve of hydrogen could still be trapped in its mantle \citep{kite2019superabundance}. This could lead to the out-gassing of a secondary atmosphere, as suggested in the case of the Earth-like planet GJ 1132 $b$ \citep{swain2021detection}. 
Precise radial velocity measurements and improved mass estimates will be critical to determining the mean density of the planet and constraining the planet's composition. TOI-411 $c$ and the single transit event candidate appear to be, respectively, a hot and a warm sub-Neptune, hence they could potentially hold a thick H/He envelope. 
Future observations with new generation telescopes of these $< 4 \ R_{\oplus }$ exoplanets, will be crucial to identify their atmospheric composition and narrow down the possibilities regarding their formation mechanisms. 
For this purpose, following the methods of \citet{Kempton2018}, we estimated the potential for atmospheric follow-up of these worlds with the James Webb Space Telescope (JWST). We computed the Transmission Spectroscopy Metric (TSM), a quantity proportional to the expected SNR of the transmission spectrum of a cloud-free atmosphere, over a 10-hour observation in the NIRISS bandpass. 
The TSM values for planets $b$ and $c$ and candidate $d$ are, respectively, $65\pm10$, $89\pm16$ and $67\pm14$. We note that the TSM value is intended as a rough indicator for transmission spectroscopy follow-ups that can and should be refined with precise mass measurement of the planets.
We put our planets into context by comparing their TSM with the same quantity for similar worlds. Specifically, Fig.\ref{fig:TSM} shows the TSM values for every confirmed planet listed on the NASA Exoplanet Archive\footnote{https://exoplanetarchive.ipac.caltech.edu/index.html} orbiting F/G stars, with $1.5 R_{\oplus} < R_{p} <4.0 R_{\oplus}$ and $P_{orb}< 50 d$. The lower limit is chosen to discard the terrestrial planets for which the \citet{Kempton2018} TSM threshold is set to 10 whilst the upper one keeps the sample within the sub-Neptune radius range. The reported equilibrium temperature ($T_{eq}$) has been computed for each planet by assuming null albedo and full day-night heat redistribution. Within this sample of 1140 planets, only  1.2\% of them have a TSM value above the threshold of 90 that \citet{Kempton2018} indicates to qualify the planet as suitable for atmospheric detection with JWST, and TOI-411 $c$ can potentially overcome this threshold.

\section{Conclusions}
\label{conc}
We present the discovery of a multi-planetary system around the bright ($V\sim$8.3 mag), nearby ($\sim$63 pc) Sun-like star HD~22946, indicated also as TOI-411, from \tess data.
The host star was observed by \tess in Sectors 3,4, 30 and 31 and two transiting planets with periods of 4.04 and 9.57 days were identified in the light curve, with an additional single event that we associate to a third transiting planet with a period of $46 \pm 4$ days. By modeling the transits, we determine the planets radii: $1.74 \pm 0.10$ $R_\oplus$ for planet $b$, $2.74 \pm 0.14$ $R_\oplus$ for planet $c$ and $3.23 \pm 0.19$ $R_\oplus$ for planet $d$. 

While the nature of the single event cannot be verified by using exclusively \tess data and follow-ups are not available yet, the innermost exoplanets have been vetted under several different aspects:
the pipelines \textsf{DAVE}, \textsf{vespa} and \textsf{triceratops} have been used to exclude false positive scenarios or determine that the false positive probability is negligible, whilst TFOP photometric data and high-resolution imaging have been used to rule out faint bound companions and foreground/background or nearby eclipsing binaries that could contaminate the \tess light curve.

Therefore, we validated the discovery of planets $b$ (or TOI-411.02) and $c$ (or TOI-411.01), while we plan to collect additional data (radial-velocity curve and photometric follow-up observations) in order to further confirm the nature of candidate planet $d$ (or TOI-411.03)

Multi-planetary systems around Sun-like stars, such as the one reported here, are ideal laboratories to test our understanding of how planets form end evolve. 
TOI-411 $b$, with an inferred mass of $\sim$ 3.5 $M_{\oplus}$, is a hot super-Earth \citep{Fulton2017,Fressin2013}, orbiting its star with a period of $\approx$ 4 days at the equilibrium temperature of 1378$\pm$36 K. Because of its size, it is one of the relatively few planets found within the Fulton gap, i.e., a range of planetary radii between 1.5 and 2$R_{\oplus}$ that is relatively less populated \citep{Fulton2017} with respect to smaller or larger radii. This bimodality in the planets' distribution is thought to be the effect of photoevaporation of volatile gases or core-powered mass loss that strips low-mass planets of their atmosphere, leaving behind bare, rocky planets, while gas giants with radii larger than 2$R_{\oplus}$ remain unaffected. 

Indeed, TOI-411 $c$ and TOI-411 $d$ have larger radii, with a radius ratio of 1.6 and 1.8 respectively. By adopting the approach described in \citet{humphrey2020predicting}, we find that between planet $b$ and $c$ it would be possible to include at least another planet, without compromising the dynamical stability of the system.  In the scenario described above, this undetected planet would have an intermediate period and a smaller radius ratio. 
However, we did not find evidence for this intermediate planet from the \tess light curve. 

With an equilibrium temperature of about 1000 K, planet $c$ is a hot sub-Neptunian, and holds particular interest because it $can$ potentially host a tick atmosphere and has a TSM compatible with atmospheric detection from JWST. In fact, even if the host star is bright, its luminosity ($\sim$ 7.3 mag in the $J$-band) is still below the saturation limit for the NIRISS single-object spectroscopy mode. An independent constraint on the mass value for this planet from radial-velocity measurement is critical to confirm the suitability of this target for atmospheric detection.
On the other side, by being a cold ($T_{eq}$=622 K) sub-Neptunian, planet $d$ would provide the opportunity of making comparative atmospheric characterization via transmission spectroscopy, if observable. In its case, the TSM estimate should be recomputed when the validation is confirmed and the period is known with better precision.

In summary, the HD~22946 multiplanetary system offers many interesting insights into planetary formation, by hosting a star that is bright and similar to our Sun, three planets with a common origin but a completely different aspect: one small planet stripped of its atmosphere (or with a secondary, accreted atmosphere), two larger planets at significantly different temperature, and probably yet-to-be-discovered intermediate ones. 

Therefore, 
this system provides an exciting opportunity for a synergy of space- and ground-based facilities such as the CHaracterising ExOplanets Satellite \citep[CHEOPS][]{Broeg2013}, HARPS and ESPRESSO to determine the bulk density of the planets, and, possibly, JWST to investigate the presence of the atmosphere. This will help us to gain a better understanding of the formation pathways that can lead to a similar system. 

\begin{acknowledgements}
We thank the referee for her/his comments that helped to improve and clarify the presentation of our results. 

This paper includes data collected by the \tess mission, which are publicly available from the Mikulski Archive for Space Telescopes (MAST). Funding for the TESS mission is provided by NASA's Science Mission Directorate.

We acknowledge the use of public \tess Alert data from pipelines at the \tess Science Office and at the \tess Science Processing Operations Center.

This research has made use of the Exoplanet Follow-up Observation Program website, which is operated by the California Institute of Technology, under contract with the National Aeronautics and Space Administration under the Exoplanet Exploration Program. 

Resources supporting this work were provided by the NASA High-End Computing (HEC) Program through the NASA Advanced Supercomputing (NAS) Division at Ames Research Center for the production of the SPOC data products.

L.I. acknowledges the “PON Ricerca e Innovazione: Attraction
and International Mobility (AIM)” program for support.

This work has made use of data from the European Space Agency (ESA) mission {\it Gaia} (\url{https://www.cosmos.esa.int/gaia}), processed by the {\it Gaia} Data Processing and Analysis Consortium (DPAC, \url{https://www.cosmos.esa.int/web/gaia/dpac/consortium}).

Observations in the paper made use of the High-Resolution Imaging instrument Zorro at Gemini-South. 

The data underlying this article is publicly available at the following archives: MAST\footnote{https://archive.stsci.edu/missions-and-data/tess} and ExoFOP-TESS\footnote{https://exofop.ipac.caltech.edu/tess/}.\\
This work is based on data collected at the following \textit{facilities}: ASAS-SN,
Exoplanet Archive,
\emph{Gaia},
Gemini:South (Zorro),
MAST,
LCOGT,
SOAR (HRcam),
TESS.\\
This work makes use of the following \textit{software}: 
 \textsf{astropy}, \textsf{lightkurve} \citep{Lightkurve2018}, \textsf{transitleastsquares} \citep{Hippke2019}, \textsf{DAVE} \citep{Kostov2019}, \textsf{exoplanet} [\citep{exoplanet:foremanmackey17},\citep{exoplanet:foremanmackey18}], \textsf{vespa} \citep{Morton2015}, \textsf{triceratops} \citep{Giacalone2020}, \textsf{tpfplotter} \citep{Lillo-Box2020}.\\

\end{acknowledgements}

\bibliographystyle{aa} 
\bibliography{biblio.bib} 

%

\begin{appendix} 
\section{DAVE results}
\begin{figure}
    \centering
    \includegraphics[width=0.5\textwidth]{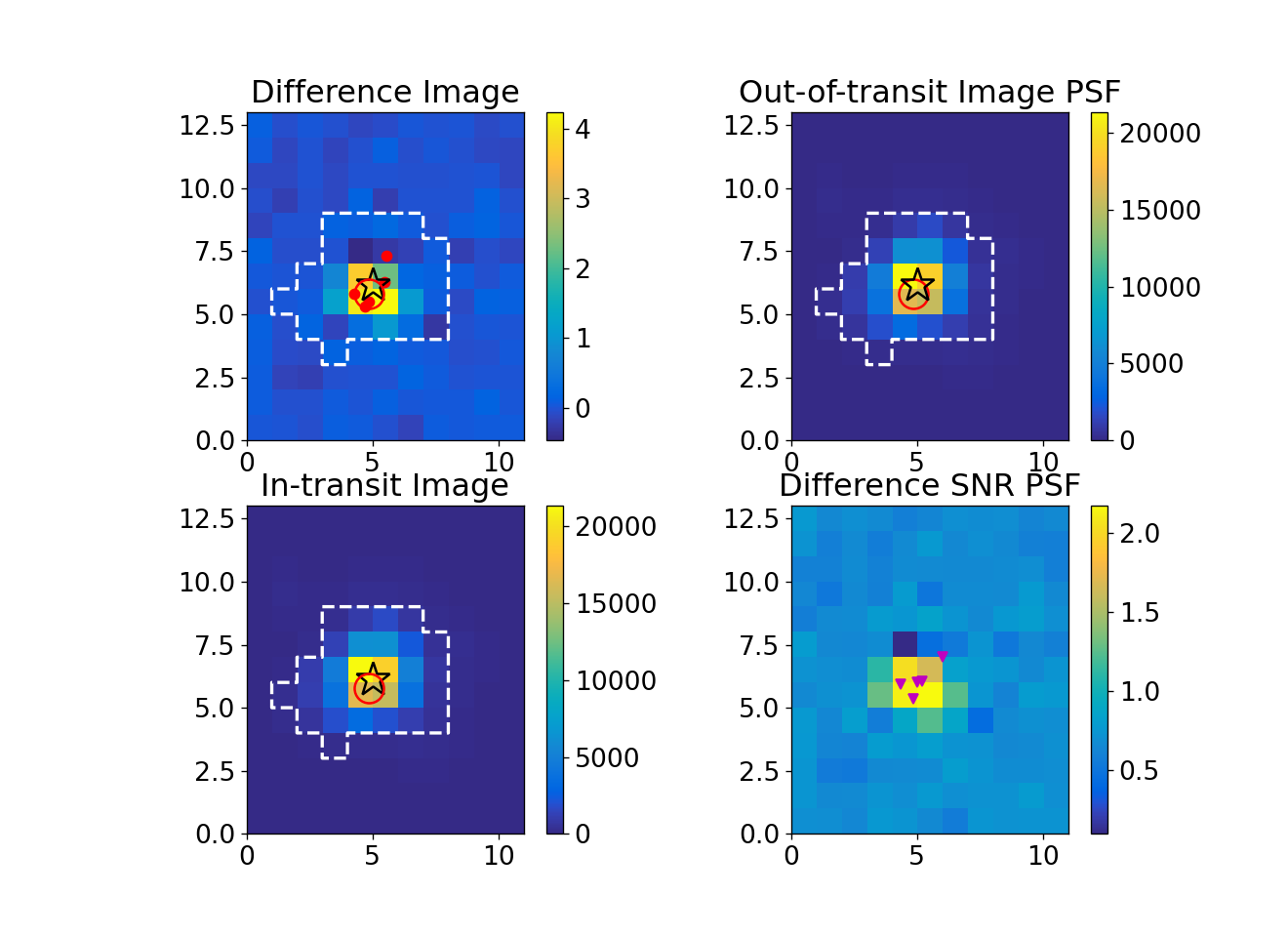} 
    \includegraphics[width=0.5\textwidth]{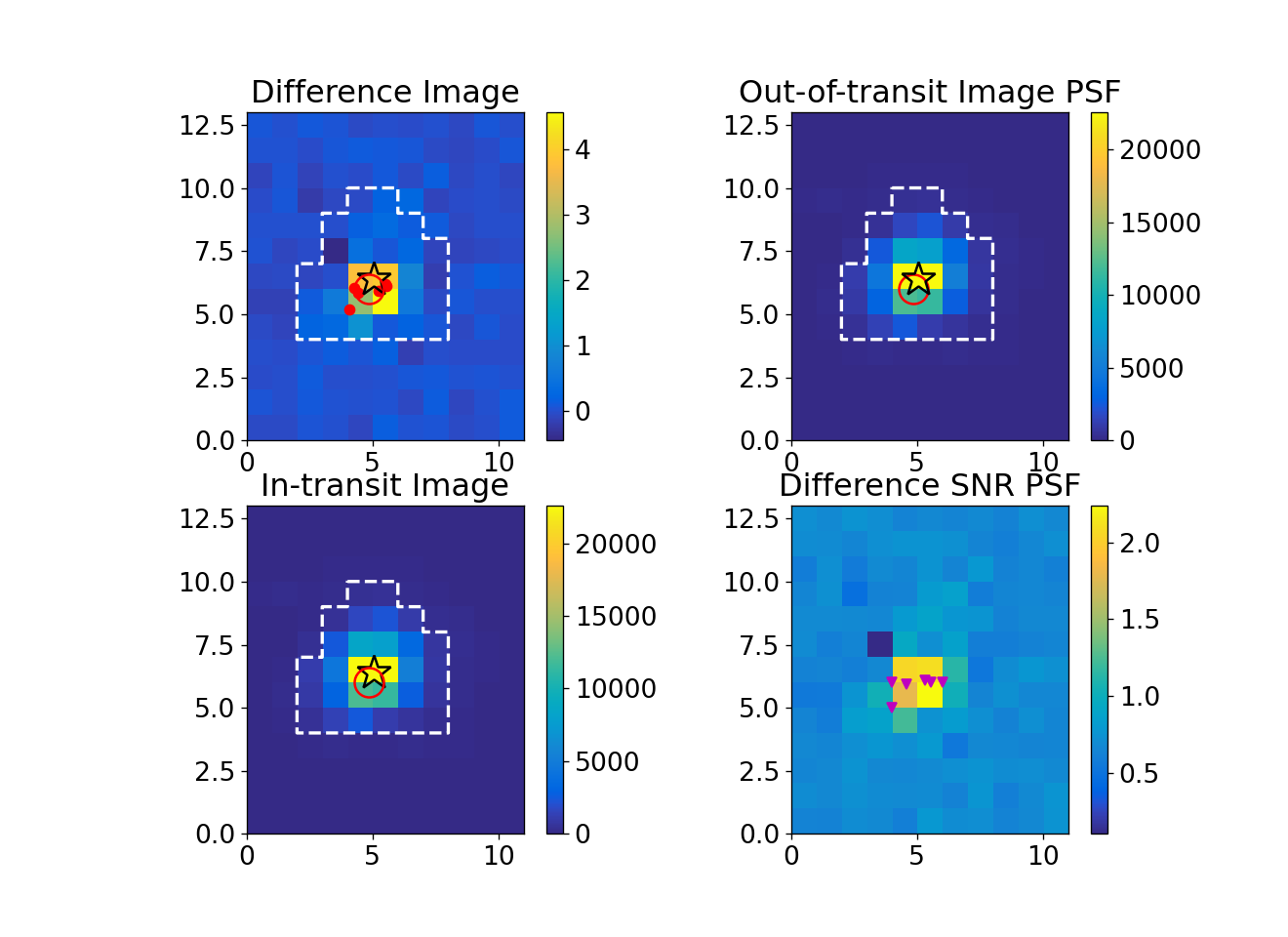} 
    \caption{\textsf{DAVE} photocenter out-of-transit, in-transit, difference and difference SNR images for planet b based on TESS sector 3 (upper panel) and 30 (upper panel) data. The axes numbers are expressed in TESS pixels.}
    \label{fig:centroidsb}
\end{figure}

\begin{figure}
    \centering
    \includegraphics[width=0.5\textwidth]{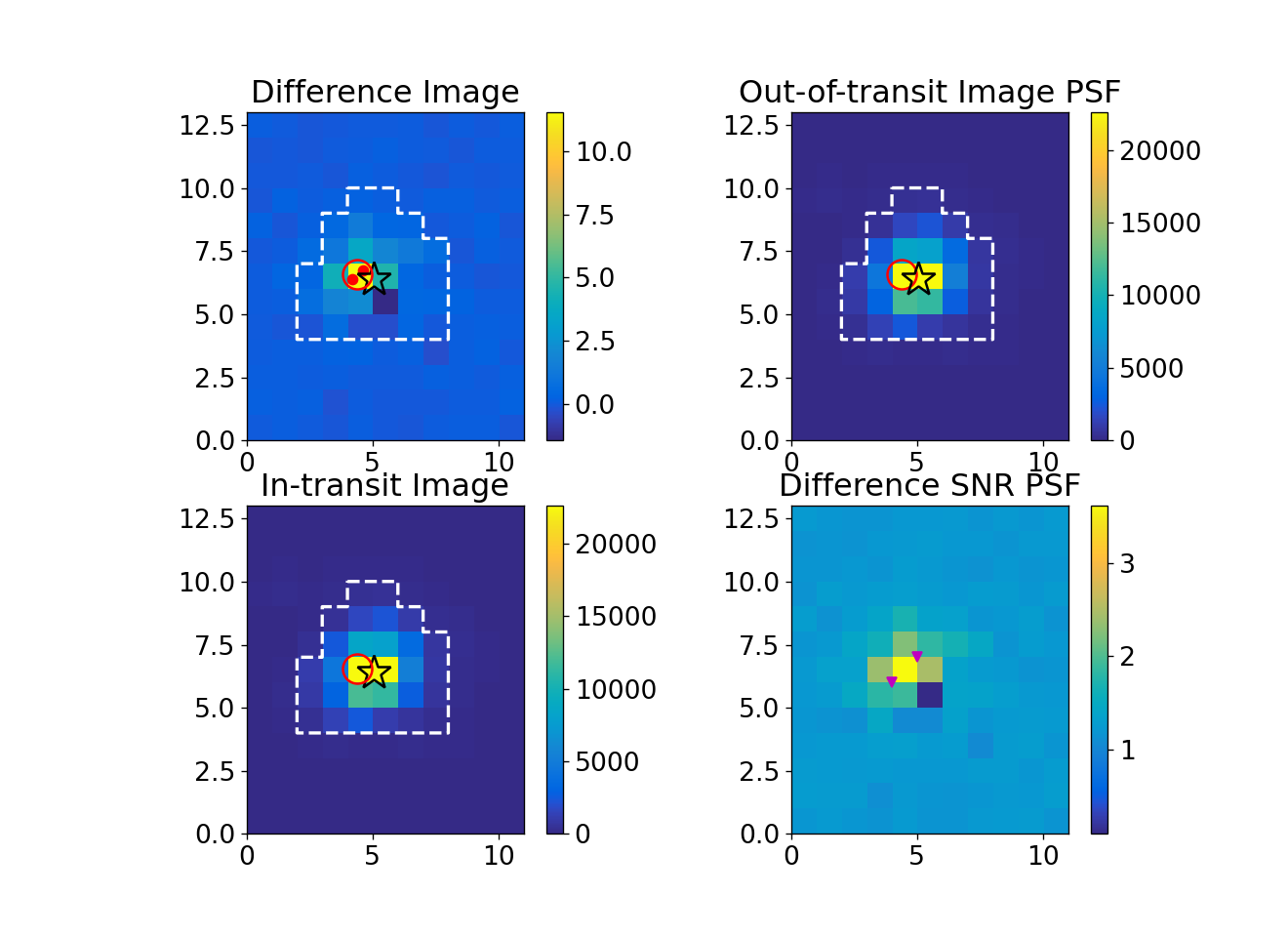} 
    \includegraphics[width=0.5\textwidth]{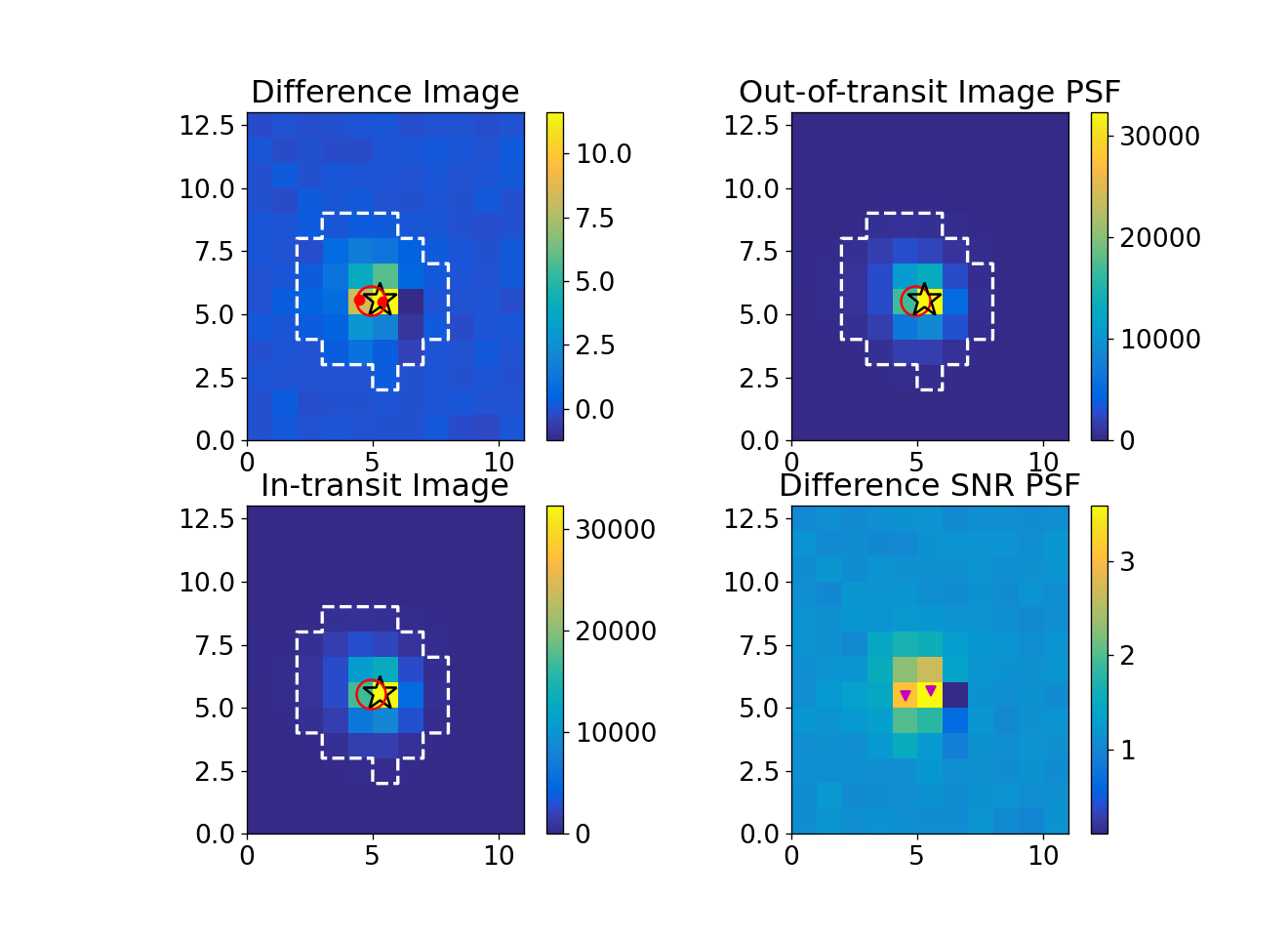} 
    \caption{\textsf{DAVE} photocenter out-of-transit, in-transit, difference and difference SNR images for planet c based on sector 30 (upper panel) and 31 (lower panel) data. The axes numbers are expressed in TESS pixels.}
    \label{fig:centroidsc}
\end{figure}

\begin{figure*}
    \centering
    \includegraphics[width=0.7\linewidth]{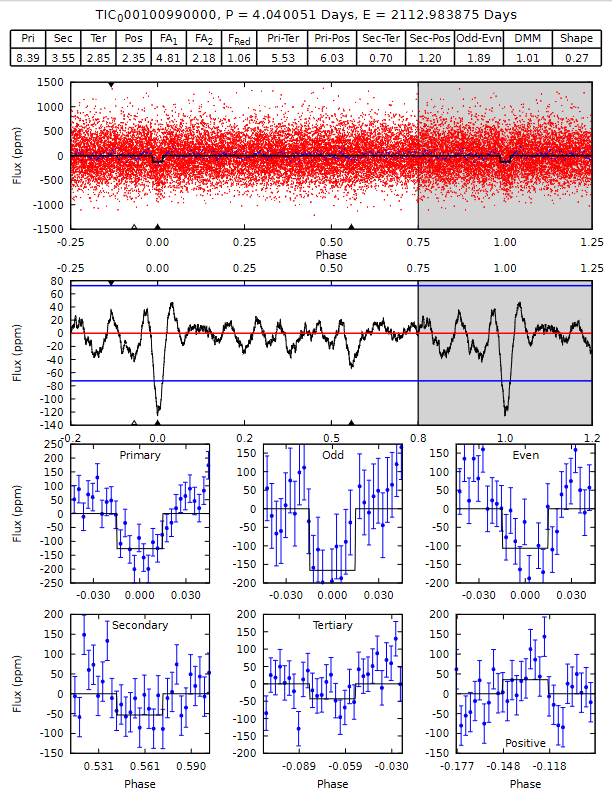}
    \caption{\textsf{DAVE} \textsf{Modelshift} for planet b based on TESS data of Sector 30. No significant false positive indicator is detected.}
    \label{fig:modshiftb}
\end{figure*}

\begin{figure*}
    \centering
    \includegraphics[width=0.7\linewidth]{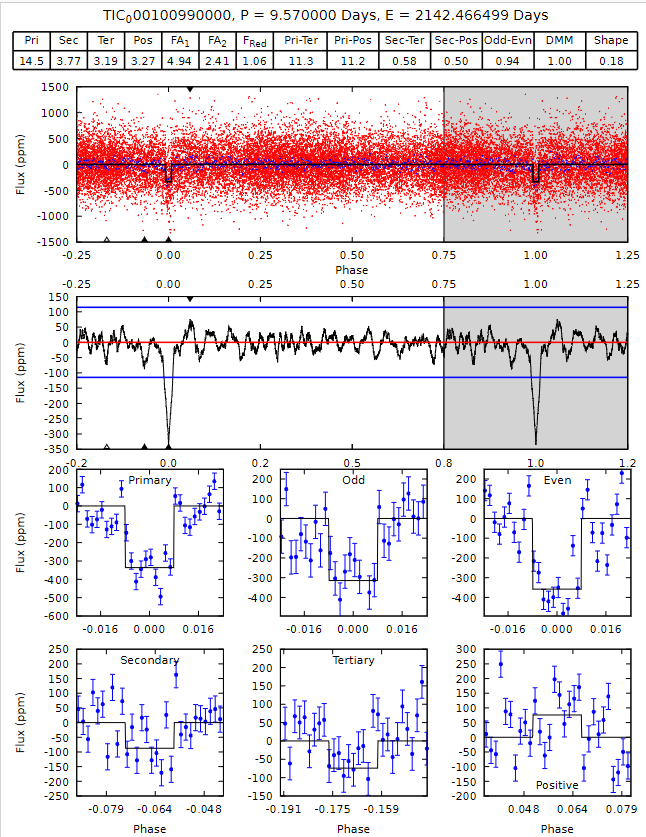}
    \caption{\textsf{DAVE} \textsf{Modelshift} for planet c based on TESS data of Sector 31. No significant false positive indicator is detected.}
    \label{fig:modshiftc}
\end{figure*}

In this Section we show examples of \textsf{DAVE} results.
Figures \ref{fig:centroidsb} and \ref{fig:centroidsc} show the photocenter difference images for the planet candidates as computed in different sectors. The black star indicates the TIC position, the red circle is the average of measured photocenters for each transit. The white dashed line indicates the TESS target pixel aperture used to extract the light curve.
Figures \ref{fig:modshiftb} and \ref{fig:modshiftc} show TESS transit data (first panel), the binned data (second panel), diagnostic plots to check for odd-even effects (third panel) secondary, tertiary eclipses and positive bumps, as might be produced by false positives (bottom panel).
DAVE analysis reports no significant false positive indicators for TOI~411 $b$ and $c$.

\FloatBarrier
 
\end{appendix}

\end{document}